\documentclass[journal]{vgtc}                          %
\ifpdf%
  \pdfoutput=1\relax                   %
  \usepackage{graphicx}                %
  \DeclareGraphicsExtensions{.pdf,.png,.jpg,.jpeg} %
\else%
  \usepackage{graphicx}                %
  \DeclareGraphicsExtensions{.eps}     %
\fi%

\graphicspath{{figures/}{pictures/}{images/}{./}} %

\usepackage{microtype}                 %
\PassOptionsToPackage{warn}{textcomp}  %
\usepackage{textcomp}                  %
\usepackage{mathptmx}                  %
\usepackage{times}                     %
\usepackage{cite}                      %
\usepackage{tabu}                      %
\usepackage{booktabs}                  %
\usepackage{subfigure}
\usepackage{amsmath,amssymb}
\usepackage[ruled,noend,noline]{algorithm2e}

\newcommand{\hedit}[1]{#1}%

\onlineid{1368}

\vgtccategory{Research}

\vgtcinsertpkg

\title{Edge-Path Bundling:  A Less Ambiguous Edge Bundling Approach}

 \author{Markus Wallinger\thanks{e-mail: \{mwallinger,  noellenburg\}@ac.tuwien.ac.at}\\ 
         \scriptsize TU Wien 
 \and 
 Daniel Archambault\thanks{e-mail: d.w.archambault@swansea.ac.uk}\\ %
      \scriptsize Swansea University
 \and David Auber\thanks{e-mail: auber@labri.fr}\\ %
      \scriptsize University of Bordeaux%
 \and Martin Nöllenburg$^*$\\
      \scriptsize TU Wien %
 \and Jaakko Peltonen\thanks{e-mail: jaakko.peltonen@tuni.fi}\\ %
      \scriptsize Tampere University }

\teaser{
  \centering
      \subfigure[Straight Line Drawing]{\label{fig:teaserSL}\includegraphics[width=0.48\linewidth]{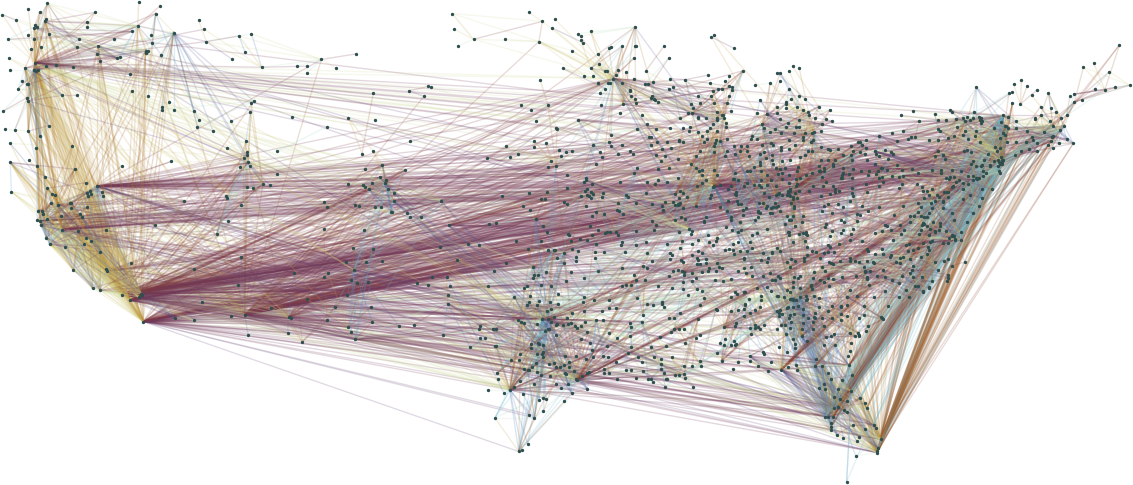}}
      \subfigure[Force-Directed]{\label{fig:teaserFB}\includegraphics[width=0.48\linewidth]{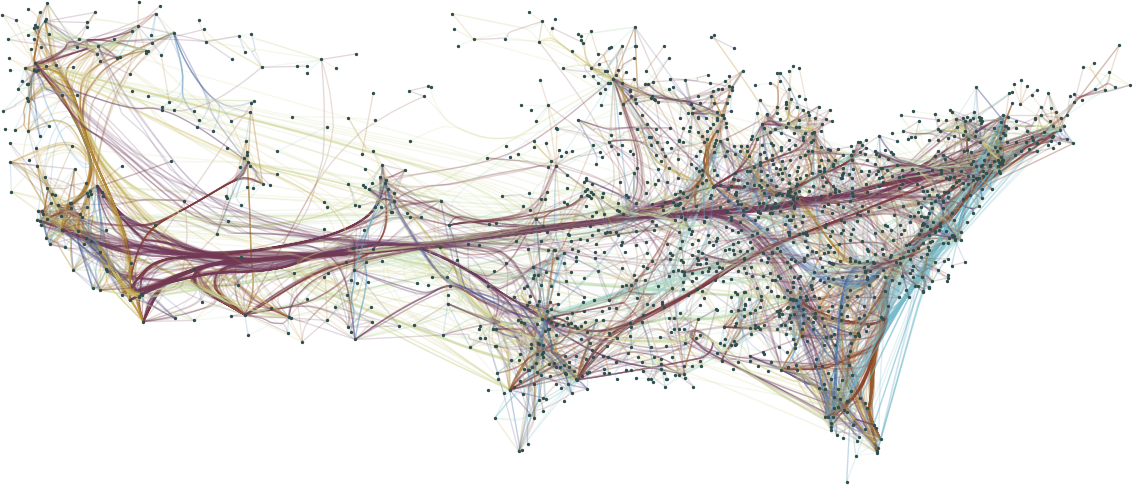}}
      \subfigure[CUBu]{\label{fig:teaserCubu}\includegraphics[width=0.48\linewidth]{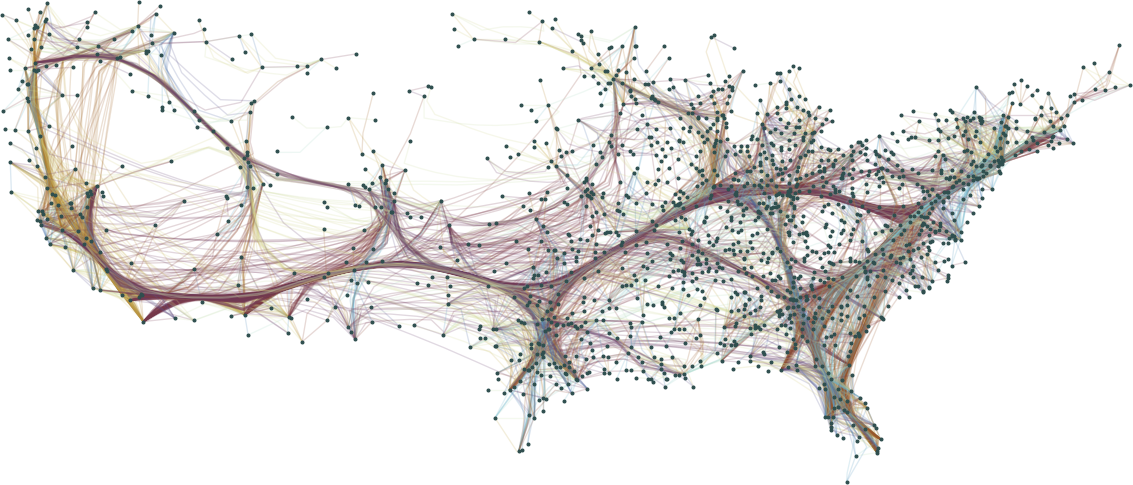}}
        \subfigure[Winding Roads]{\label{fig:teaserWR}\includegraphics[width=0.48\linewidth]{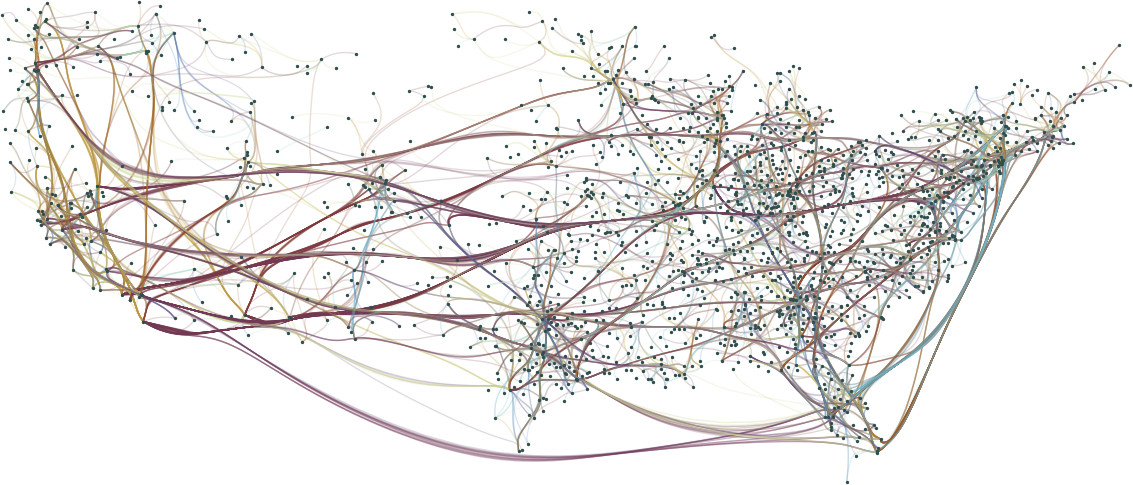}}
      \subfigure[Edge-Path Bundling]{\label{fig:teaserEPB}\includegraphics[width=0.48\linewidth]{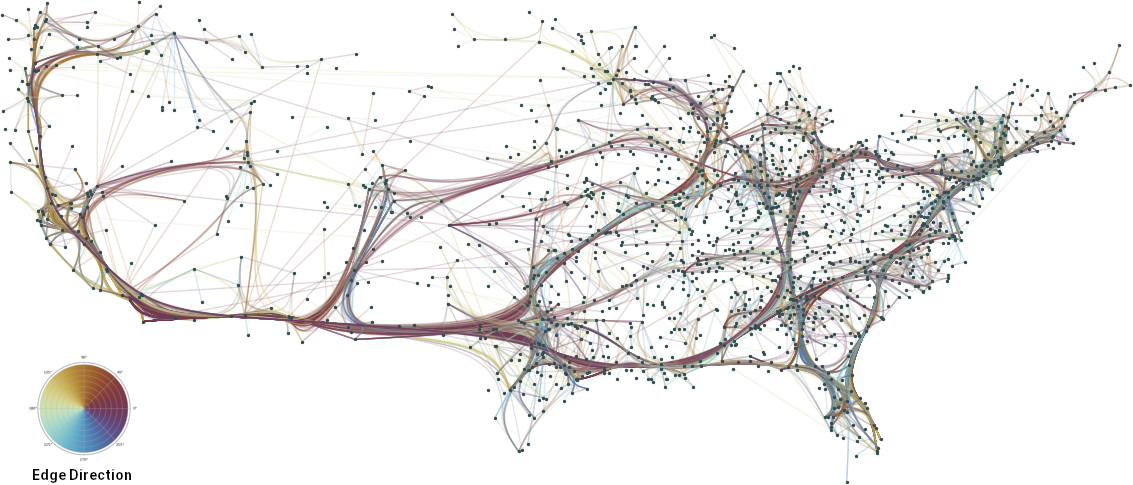}}
      \subfigure[Edge-Path Bundling Directed]{\label{fig:teaserEPBDir}\includegraphics[width=0.48\linewidth]{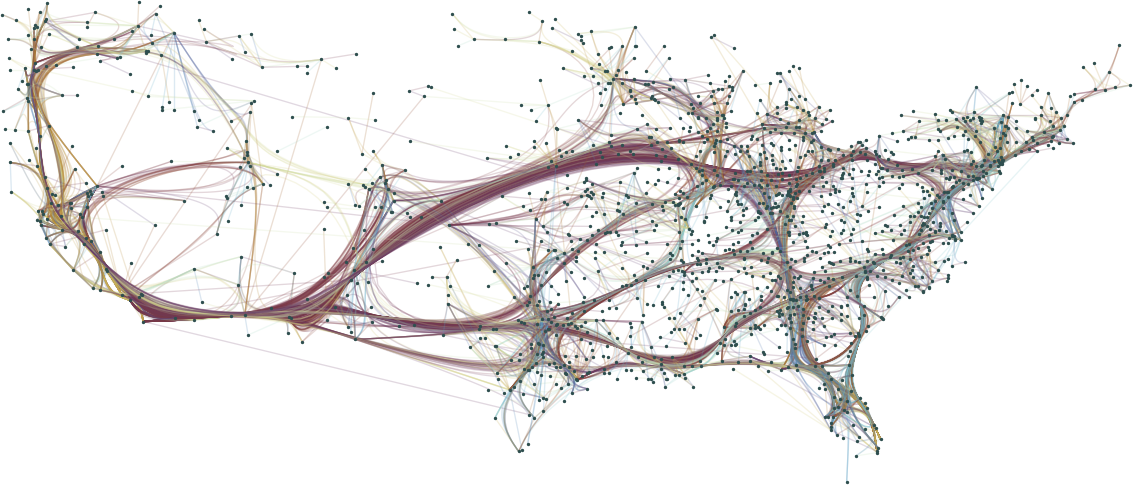}}
  \caption{Edge bundling of the Migrations dataset.  {\bf (a)} Straight line drawing.  {\bf (b)} Force-Directed edge bundling aggregates edges well, but overaggregates at the centre of the drawing making it difficult to see patterns in the east-west flow (red bundle). {\bf (c)} CUBu has a similar drawback at the centre of the map to a lesser degree.  {\bf (d)} Winding Roads divides this structure into several smaller flows, but they may not be necessarily related to graph structure.  {\bf (e)} Edge-Path bundling is able to distinguish between several flows that reflect paths in the underlying graph.  {\bf (f)} When edge direction is considered, the algorithm is able to further subdivide these flows based on direction.}
  \label{fig:teaser}
}

\abstract{%
Edge bundling techniques cluster edges with similar attributes (i.e. similarity in direction and proximity) together to reduce the visual clutter.  All edge bundling techniques to date implicitly or explicitly cluster groups of individual edges, or parts of them, together based on these attributes.  These clusters can result in ambiguous connections that do not exist in the data.  Confluent drawings of networks do not have these ambiguities, but require the layout to be computed as part of the bundling process.  We devise a new bundling method, Edge-Path bundling, to simplify edge clutter while greatly reducing ambiguities compared to previous bundling techniques.  Edge-Path bundling takes a layout as input and clusters each edge along a weighted, shortest path to limit its deviation from a straight line.  Edge-Path bundling does not incur independent edge ambiguities typically seen in all edge bundling methods, and the level of bundling can be tuned through shortest path distances, Euclidean distances, and combinations of the two.  Also, directed edge bundling naturally emerges from the model.  \hedit{Through metric evaluations, we demonstrate the advantages of Edge-Path bundling over other techniques. } %
} %

\CCScatlist{ 
}

\begin{document}

\maketitle
\section{Introduction} %
\begin{figure}
    \centering
    \subfigure[]{\label{fig:ambcase}\includegraphics[width=0.125\linewidth]{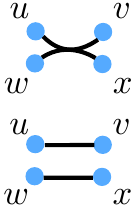}}
    \hspace{0.5cm}
    \subfigure[]{\label{fig:undiredgepath}\includegraphics[width=0.35\linewidth]{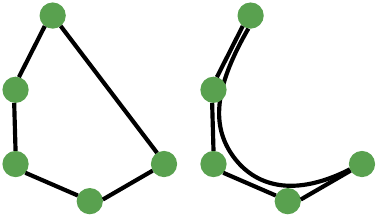}}
    \hspace{0.5cm}
    \subfigure[]{\label{fig:diredgepath}\includegraphics[width=0.35\linewidth]{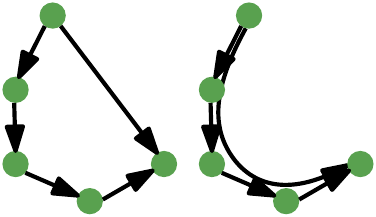}}
    \caption{Independent edge ambiguity and Edge-Path bundling. {\bf (a)} Two disconnected edges can be bundled together in edge bundling approaches resulting in false connections between $(u, x)$ and $(v,w)$. {\bf (b)} Edge-Path bundling avoids this issue by bundling long edges with weighted shortest paths. {\bf (c)} For directed graphs, directed paths are used.}
    \label{fig:bundleTypes}
\end{figure}

Since its introduction~\cite{holten_2006}, edge bundling approaches cluster groups of individual edges together based on similar properties.  The original paper required a hierarchy 
imposed 
on top of the network, but quickly 
techniques were developed to bundle edges with shared attributes~\cite{lhuillier_2017} (e.g.~proximity and movement in the same direction~\cite{cui_2008,holten_2009,lambert_2010,telea_2010}) (see Fig.~\ref{fig:teaser}). However, all 
these approaches implicitly or explicitly cluster individual edges, or parts of them (such as their pixels), independent of graph structure. This causes a 
type of ambiguity shown in Fig.~\ref{fig:ambcase} 
called the {\it independent edge ambiguity}: two independent edges can be clustered together leading to the perception of false adjacencies that do not exist in the underlying graph.

Confluent drawings~\cite{dickerson_2005} considered a very similar problem, and cluster edges based on their participation in bicliques.  Therefore, confluent drawings do not suffer from independent edge ambiguities.  However, confluent drawings bundle graphs significantly less, leaving much edge clutter on-screen.  Also, these approaches compute both the layout of the network and the bundling simultaneously.  There have been a number of attempts to relax the strict constraints of confluent drawings to move towards unambiguous bundling~\cite{bach_2016,zheng_2021}, but all approaches still require the layout to be computed alongside the bundling.

In all such approaches, groups of edges with similar attributes (typically, edge slope and proximity) or parts of them (i.e. their pixels) are bundled together if they are aligned well.  However, when clustering these edges, the underlying graph structure is not fully considered.  In Fig.~\ref{fig:teaser}, a straight line drawing and three popular bundling approaches are shown.  Force-Directed edge bundling~\cite{holten_2009} and CUBu~\cite{vdzwan_2016} greatly simplify the edge clutter clearly revealing the direction of flows.  This is particularly visible in the east-west flows of the network (the red edges) at the centre of the drawing.  However, unrelated edges can be pulled together into a bundle causing ambiguity in the patterns of connections.  Grid-based techniques, such as Winding Roads~\cite{lambert_2010}, divide the edges into much smaller bundles, but these bundles can contain unrelated edges.  The approach presented in this paper, Edge-Path bundling, bundles edges with shortest paths between their endpoints.  Therefore, unrelated edges will not be bundled and all bundles reflect an underlying path in the graph.  In the undirected version, it is now clear that the central flow divides into two: one that heads towards the great lakes region and another towards Texas.  Further detail is also visible on the east coast and the great lakes region not visible in the other diagrams.  Directed Edge-Path bundling reveals that these bundles actually consist of three main streams indicated in red.  In the straight line drawing, these flows are somewhat visible, but the pattern is not revealed by any other technique except directed Edge-Path bundling; in particular, the flow from California to Texas divides into three.

This paper introduces a new approach to edge bundling that does not consider groups of individual edges, their pixels, or bicliques as the primitive for bundling.  Instead, given an input layout (i.e. trail-sets~\cite{lhuillier_2017}), it considers clustering edges with a weighted path as the primitive for bundling.  Each long edge in the graph is bundled to a shortest path that exists between the endpoints of the edge (Fig.~\ref{fig:undiredgepath} and \ref{fig:diredgepath}).  By definition, Edge-Path bundling does not suffer from independent edge ambiguities as a path must exist in the graph for the edge to be bundled, but it is far less restrictive than the rules imposed by confluent drawings.  Also, the approach naturally expresses both undirected and directed edge bundling without modification to the algorithm.  We demonstrate that the approach has more significant bundling when compared to confluent drawings while simultaneously eliminating independent edge ambiguities.

\section{Related Work}

Since the introduction of edge bundling by Holten~\cite{holten_2006} and confluent drawing by Dickerson {\it et al.}~\cite{dickerson_2005} for reducing edge clutter in graph drawings, the research area has been very active and many approaches have been devised by clustering edges, or their pixels, with similar properties together.  The current state-of-the-art in edge bundling can incur independent edge ambiguities (Fig.~\ref{fig:ambcase}) when aggregating edges into clusters.  Confluent drawings restrict the bundling to perfect bi-cliques to avoid this case, but often the imposed constraints are too strict for significant bundling to occur.

\begin{figure*}[t]
  \centering
      \subfigure[Edge-Path, Confluent]{\label{fig:noiseSL}\includegraphics[width=0.20\linewidth]{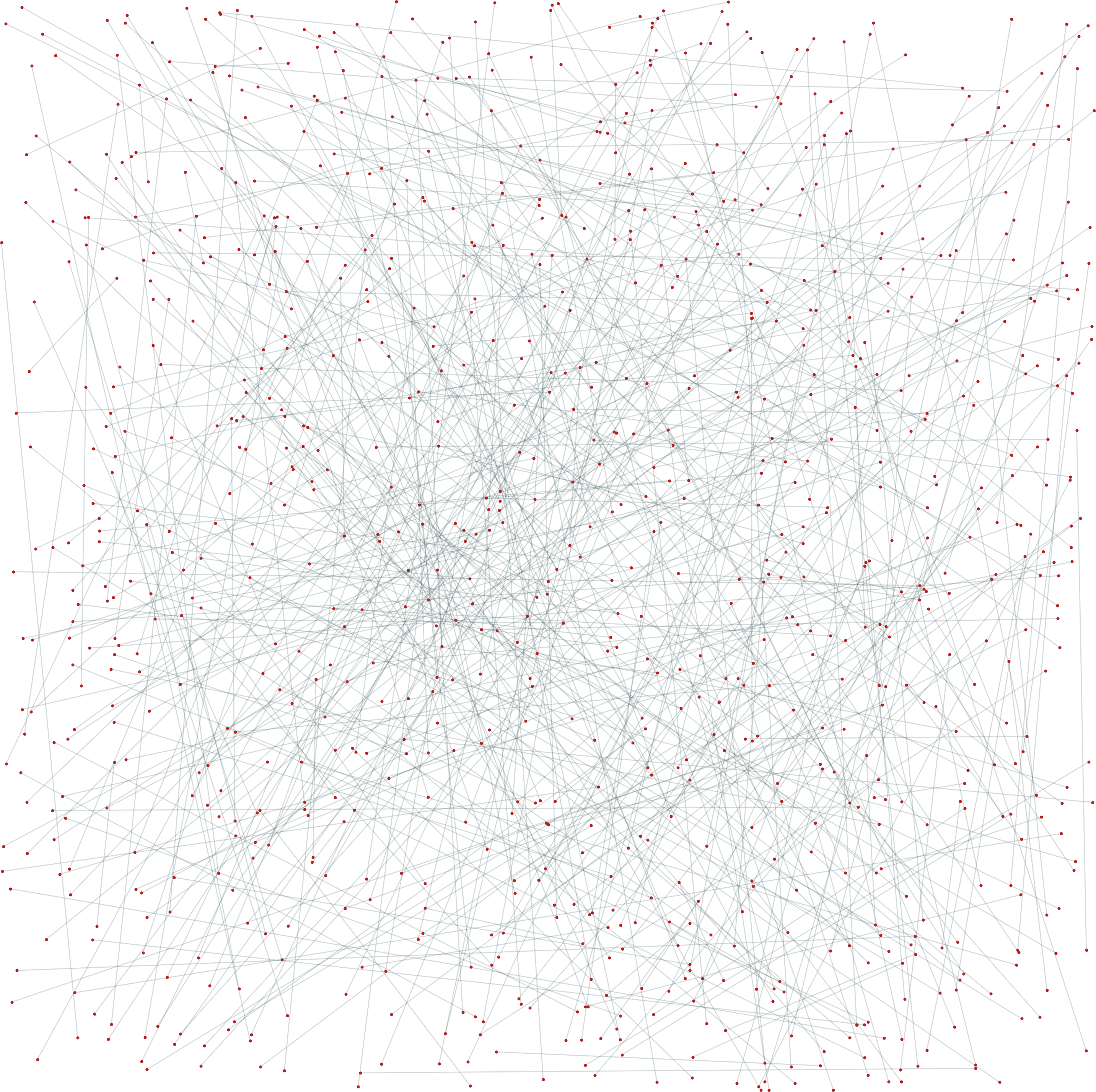}}
      \subfigure[Force-directed]{\label{fig:noiseEPB}\includegraphics[width=0.20\linewidth]{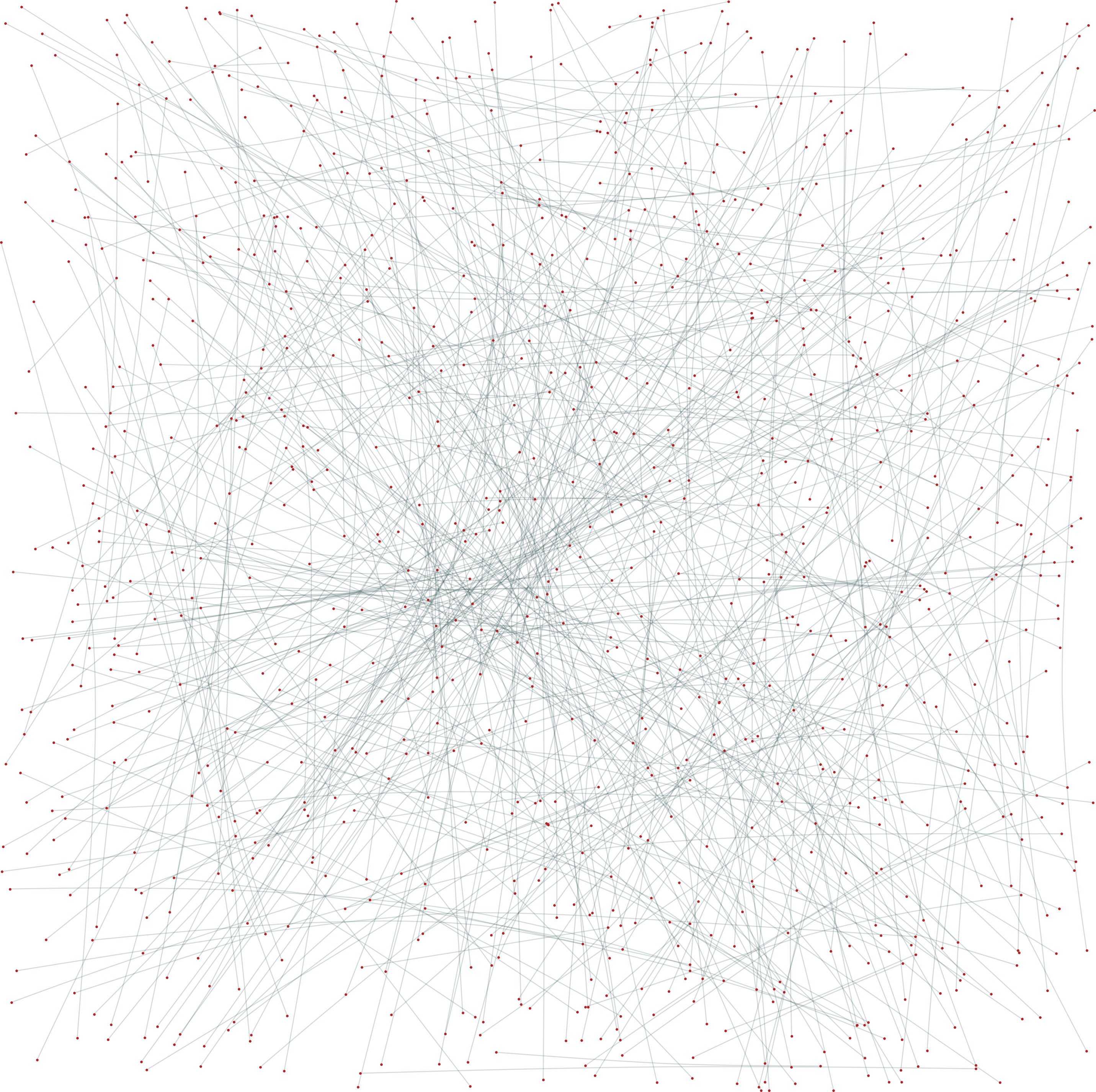}}
      \subfigure[Winding Roads]{\label{fig:noiseWR}\includegraphics[width=0.20\linewidth]{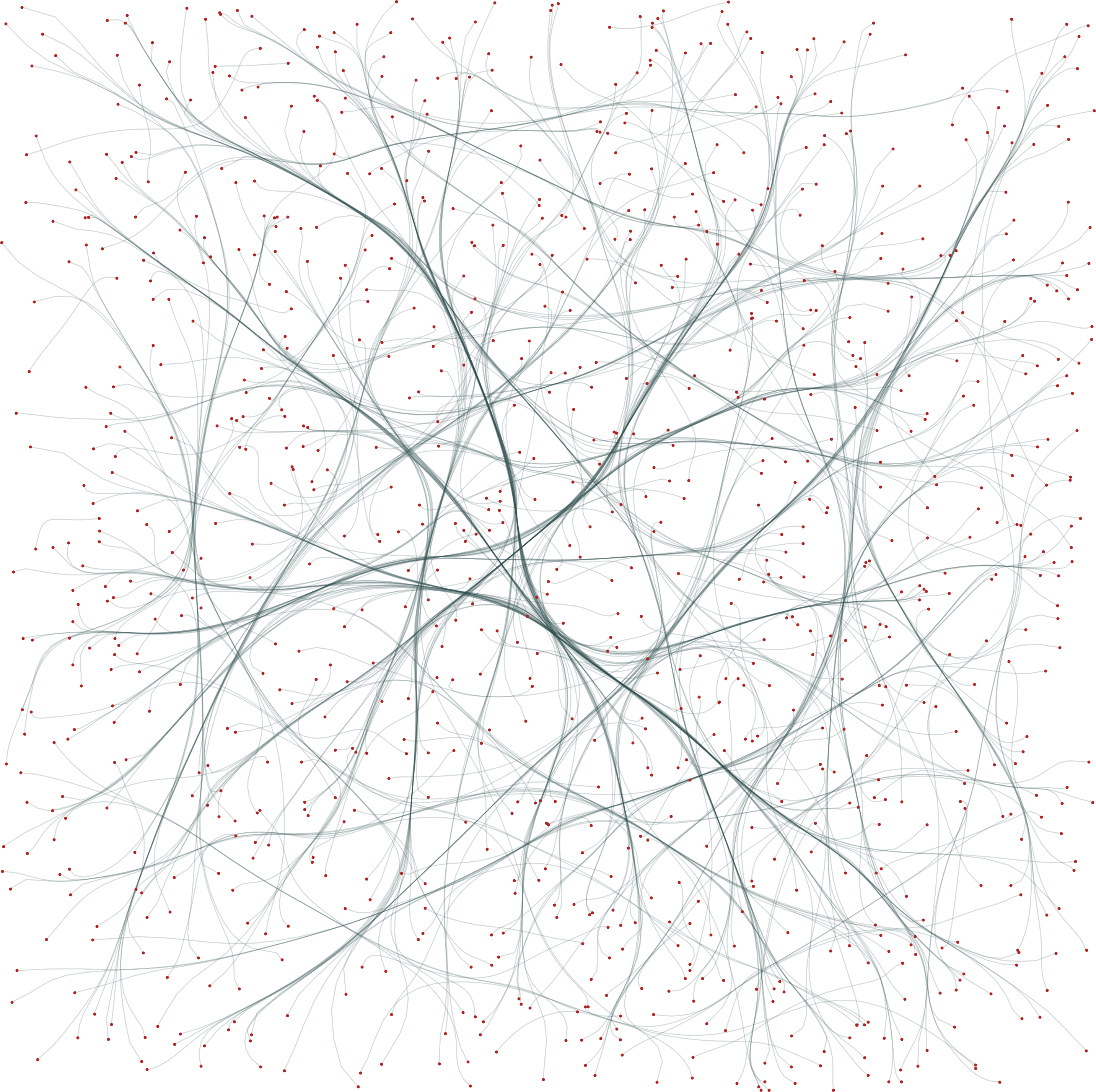}}
      \subfigure[CUBu]{\label{fig:noiseCubu}\includegraphics[width=0.20\linewidth]{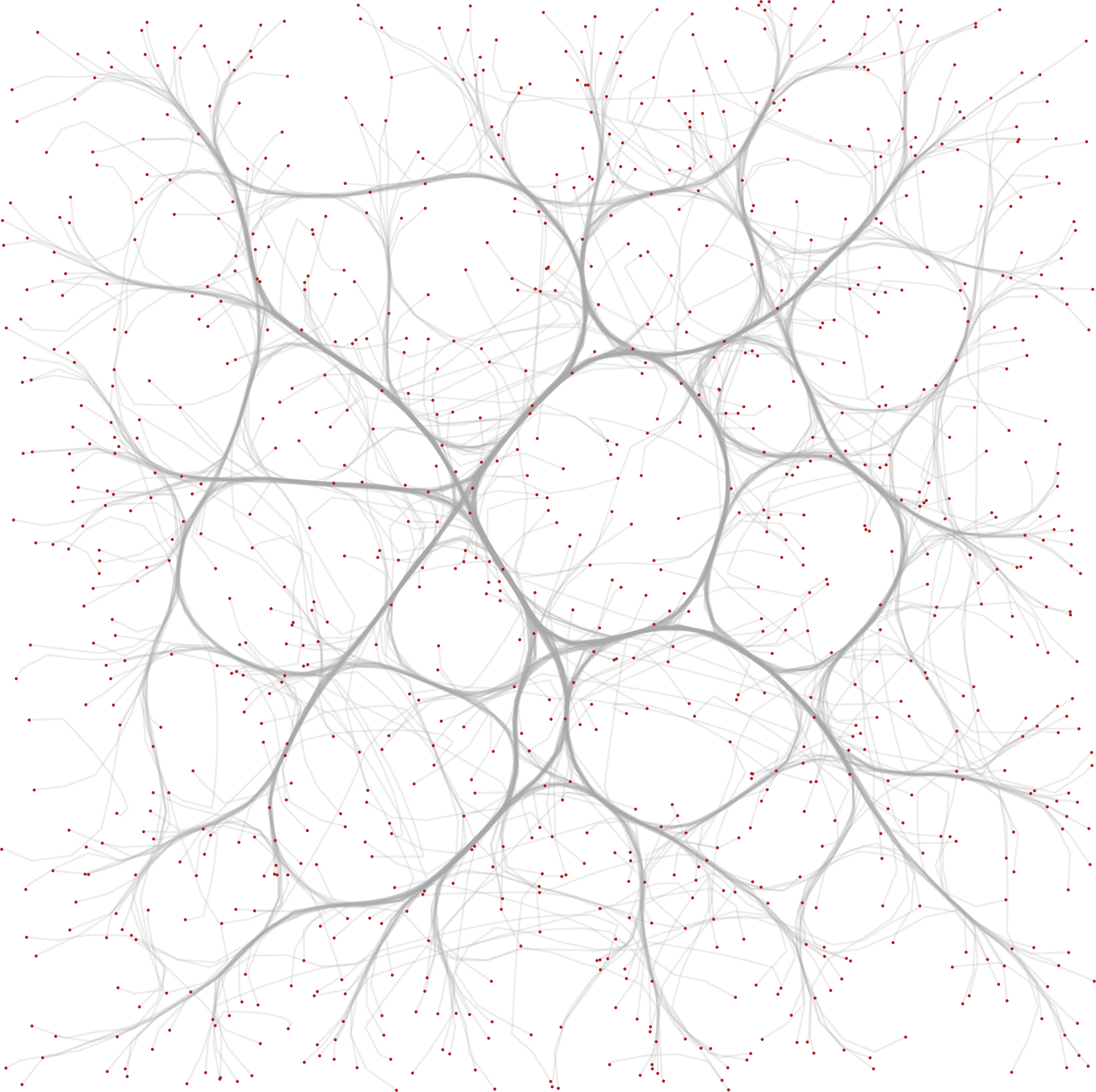}}
  \caption{Bundling of a noise graph.  Randomly placed vertices ($|V| = 1000$) are connected by a perfect matching, resulting in $|E| = |V|/2$ disconnected edges. No structure is present in this graph and there should be no bundles (See~\cite{lhuillier_2017} Figure~19).  Force-directed, Winding Roads and CUBu give the impression of structure in the underlying data when there is none.  Edge-Path bundling and confluent drawings do not find bundles.}
  \label{fig:noise}
\end{figure*}

Edge-Path bundling devises a necessary rule to have efficient bundling while completely eliminating independent edge ambiguities.  Intuitively, when a pair of disconnected edges $(u,v)$ and $(w,x)$ are bundled, a path can appear to exist between $(u, x)$ and $(v, w)$ where a connection does not exist at all.  Also, Edge-Path bundling can avoid bundling patterns where there are none.  In Fig.~\ref{fig:noise}, independent edges are placed randomly in the cube and all bundling techniques find a pattern in this graph whereas Edge-Path bundling and confluent drawings do not.  Edge-Path bundling does not have this issue as it only bundles edges with weighted paths in a particular layout. Although related to edge bundling, it is a fundamentally different approach that does not neatly fit into any of these categories.  %

{\bf Edge Bundling.} Since the first techniques were described, many bundling techniques have been explored~\cite{lhuillier_2017} often inspired by the work on flow maps~\cite{phan_2005}.  A number of techniques have been proposed, but all techniques have one common primitive:  clustering groups of edges together that share similar attributes.  

Edge bundling was introduced to the visualisation community by Holten~\cite{holten_2006}. In his seminal work, a hierarchy was superimposed on top of the network, usually via a treemap variant~\cite{johnson_1991,bruls_2000}, and the centroids of the cluster hierarchy are used as control points to cluster edges into merging and splitting streams.  Soon after, the requirement for a hierarchy was eliminated and replaced by a desired property whereby nearby edges headed in a similar direction with similar length are bundled together.  A number of approaches were created with this idea including using a triangular mesh~\cite{cui_2008}, grids and quad-trees~\cite{luo_2012,lambert_2010}, force-directed algorithms~\cite{holten_2009,nguyen_2012}, force-directed algorithms with edge direction encoded in the force system with compatibility measures (such as connection distance)~\cite{selassie_2011}, sparse visibility spanners~\cite{puprev_2016}, and multilevel clustering~\cite{gansner_2011}.  Domain-specific clusters and layouts have been used to help with the bundling process~\cite{lambert_2011,reniers_2014} as well as more general clusters~\cite{toeda_2017}. Image-based~\cite{telea_2010,hurter_2012,vdzwan_2016,ersoy_2011,wu_2015,lhuiller_2017} techniques operate on the pixels of individual edges.  These approaches create a density or similarity map computed at pixel level by summing up the contributions of all edges, after which all edges are independently advected upstream along gradients of the map. %
Bundling has also been used to simplify clusters in parallel coordinate plots~\cite{palmas_2014}.

Edge bundling techniques greatly reduce edge clutter by implicitly or explicitly clustering groups of edges or parts of them (i.e. their pixels) and \hedit{allow} 
visualisation of higher-level flow patterns in the network that otherwise would not be visible.  However, \hedit{all} 
these approaches will suffer from the independent edge ambiguities to a 
\hedit{degree}.  Also, as stated in the survey~\cite{lhuillier_2017}, random patterns can be created in data where there are no patterns.  In this paper, we seek a compromise:  efficient bundling that greatly simplifies the network while completely eliminating independent edge ambiguities.  We accomplish this by bundling long edges with a weighted shortest path between their endpoints instead of clustering groups of edges together.

{\bf Confluent Drawings.} Confluent drawings are visually similar to edge bundling but, by design, they do not suffer from independent edge ambiguities. The key idea of a confluent bundle is that only bicliques can be bundled, i.e., $K_{n,m}$ subgraphs. This guarantees that all connections implied by a bundle are actually present in the graph. In their original and theoretically motivated definition, confluent drawings do not permit any edge crossings~\cite{dickerson_2005,hpss-ttcd-07} and as a consequence some graphs do not have a confluent drawing at all. Variations of confluent drawings, such as $\Delta$-confluent~\cite{egm-dd-06} and strict confluent drawings~\cite{fgkn-sog-19,ehlnsv-scd-16}, have been studied from a theoretical point of view, showing mathematical characterizations and the NP-completeness of recognizing graphs that admit certain types of confluent drawings. While most results on confluent drawings do not provide algorithms or implementations, Dickerson \textit{et al.}~\cite{dickerson_2005} and Hirsch \textit{et al.}~\cite{hmr-becgcd-07} present and implement some heuristics based on detecting cliques and bicliques in order to introduce confluent bundles. 
Confluent drawings are based on an abstract graph input and typically solve both the graph layout and the edge bundling in a combined way. Therefore, existing confluent drawing techniques do not readily apply to edge bundling of pre-embedded networks.  In this paper, we devise a compromise that is not as strict as confluent layout, but more effectively bundles the network so that higher-level features become visible while reducing topological ambiguities.

{\bf Hybrid Approaches.} A number of papers have used confluent drawing approaches as a basis and have relaxed some of the more restrictive constraints of the approach to achieve greater simplification.  Bach {\it et al.}~\cite{bach_2016} explore ways to produce less ambiguous bundling by relaxing the strict planarity constraint of confluent drawings.  Their approach computes a power graph decomposition and uses the resulting hierarchy to both draw and bundle the network.  Zheng \textit{et al.}~\cite{zheng_2021} extend Bach {\it et al.}~\cite{bach_2016} to create strict, power-confluent drawings, based on additional constraints that further reduce crossings and ambiguities.

The two approaches described above were the first attempts to relax some of the constraints of confluent drawings to produce less ambiguous edge bundlings.  In this paper, we introduce a new approach with a different primitive, but that finds a similar compromise:  greater bundling while completely avoiding independent edge ambiguities.

{\bf Visualisations and Measures of Bundling Quality.} Alongside bundling algorithms, significant work has been undertaken to devise metrics to evaluate how well bundling algorithms perform with respect to each other and in general.  Metrics have been devised to measure faithfulness~\cite{NguyenEH17,neh_ofgv_2013}, entropy~\cite{WuZhuLiuYu18}, geodesic path tendency or distortion from the straight line distance~\cite{heh-grbgt-09}, and data-ink ratio to quantify simplification~\cite{t-vdqi-83}.  We use and adapt these metrics in our evaluation.  

Wang {\it et al.}~\cite{wang_2016} 
\hedit{give}
methods 
\hedit{to visualise}
ambiguities in graph layouts, including bundling.  In this approach, alignment in edge direction and proximity, a classic definition for bundling suitability, are used to highlight ambiguities in the graph.  Nguyen {\it et al.} define the notion of faithfulness in graph visualisation: {\it ``the underlying network data and the visual representation are logically consistent''}~\cite{neh_ofgv_2013}.  Edge bundling ambiguities is used to illustrate 
\hedit{faithfulness} or lack of 
\hedit{it}
in a representation.
Our work, by definition, increases the faithfulness of bundling representation by avoiding certain types of ambiguity.   

{\bf Summary.}  Edge bundling techniques improve %
\hedit{graph readability}, but suffer from independent edge ambiguities and can introduce patterns where none exist in the underlying data. Confluent drawings and hybrid approaches do not suffer from independent edge ambiguities, but have a lower degree of bundling and compute the drawing of the network as part of the bundling process.  
 Thus, this paper introduces a new bundling primitive (edge to path) and a new bundling algorithm based on this primitive, that produces a more faithful bundling~\cite{neh_ofgv_2013} of a layout.%

\section {Edge-Path Bundling Algorithm}\label{sec:algorithm}

Algorithm~\ref{alg:pathbundle} presents the pseudocode for our approach.  The method is surprisingly simple and  requires only four parameters:  the network $G = (V, E)$, a drawing $D_G$ of $G$, a maximum distortion threshold $k$, and an edge weight factor $d$.  The algorithm creates a local hashset, {\it lock}, that indicates when edges will be excluded from bundling by the algorithm. It also uses a second hashset {\it skip} which includes edges that will be skipped from shortest path calculations.   Initially, all edges in both hashsets are false.  The algorithm takes into account the length of an edge in $D_G$ to determine its suitability for inclusion in a shortest path: shortest paths that result in shorter detours in Euclidean space are preferred.  Therefore, a third hashset stores the weight of each edge which is the Euclidean edge length raised to a power $d$.  The exponent $d$ can be used to tune exclusion of short edges from the bundling.

As a first step, the edges are sorted in decreasing weight order and then processed in that order.  Thus, our algorithm prefers to bundle long edges over short ones.  If an edge is locked, the algorithm does not process it.  Otherwise, it is excluded from shortest path computations and processed.  The first stage of bundling is determining the shortest, weighted path between the endpoints of the edge, excluding the edge itself and previously bundled edges from this calculation.  In order to compute this path, we use Dijkstra's algorithm to compute the shortest path that takes the minimum detour from straight line distance.  If such a path does not exist or the detour is greater than a distortion threshold of $k$ times the straight line distance, the edge is not bundled and it is reintroduced into shortest path calculations.  Otherwise, the edge is bundled and the algorithm uses the vertices along the path in the drawing as control points for the edge.  The bundled edge is excluded from all future shortest path calculations as it will be rendered using a curve.  As all the edges along the path are now included in an Edge-Path bundle, these edges are locked.  Edges along the path should not be bundled as they serve as the control points for one or more bundled edges in the graph, but they still can participate in shortest path calculations to bundle other edges. \hedit{The final drawing of the network uses these control points to render each edge using a smooth B\'ezier curve. A smoothing parameter allows the algorithm to control the bundling strength in the final drawing by inserting additional control points along the path. A factor of one places control points on the vertices. Increasing this factor to two places additional control points on the centre of the straight line between two consecutive control points.  Higher factors apply this rule recursively.}

Overall, Edge-Path bundling has a worst case time complexity of $O(|E|^2\log |V|)$ as the Dijkstra algorithm \hedit{(priority queue heap implementation)} runs $O(|E|)$ times.  However, the distortion threshold $k$ can be used to stop Dijkstra's algorithm once this threshold is exceeded, reducing the chance that this worst case complexity is observed.  %

\begin{algorithm}[t]
\DontPrintSemicolon
 \caption{Edge-Path Bundling Algorithm}
    \SetKwInOut{Input}{Input}
    \SetKwInOut{Output}{Output}
    \SetKwComment{Comment}{//}{}
    \Input{Graph $G = (V, E)$, input drawing $D_G$, maximum distortion $k$, edge weight factor $d$.}
    \Output{Control points for an Edge-Path bundled drawing $\Gamma$.}
    \For{$e \in E$}{
        lock($e$) $\leftarrow$ False\;
        skip($e$) $\leftarrow$ False\;
        weight($e$) $\leftarrow$ $(D_G.\text{edgeLength} (e))^d$\;
    }

    sortedEdges $\leftarrow$ sortDescending ($E$, weight)\;
    \For{$e \in$ sortedEdges} {
        \If {lock(e)} {
            {\bf continue}\;
        }
        
        skip($e$) $\leftarrow$ True\;
        $s$ $\leftarrow$ source(e)\;
        $t$ $\leftarrow$ target(e)\;
        \Comment {Dijkstra excluding edges in skip from $G$}
        $p$ $\leftarrow$ dijkstraAlgorithm ($G$, $s$, $t$, weight, skip)\;
        \If {$p$ == null} {
            skip($e$) $\leftarrow$ False\;
            {\bf continue}\;
        }
        
        \If {$p.\text{length()} > k*D_G.\text{edgeLength} (e)$} {
            skip($e$) $\leftarrow$ False\;
            {\bf continue}\;
        }
        
        \For{$m \in p$} {
            lock($m$) $\leftarrow$ True\;
        }
        controlPoints($e$) $\leftarrow p.$getVertexCoords()\;
    }
    \Return controlPoints\;
    \label{alg:pathbundle}
\end{algorithm}

\subsection{The Lesser Ambiguities of Edge-Path Bundling}

Although our approach is free of independent edge ambiguities by definition, it is not completely free of ambiguities.  We now describe the ambiguities incurred by Edge-Path bundling.  It is important to note that the two ambiguities we define here can occur in all other edge bundling methods, but have not been concretely defined previously.

\begin{figure}
\centering
\includegraphics[width=0.65\linewidth]{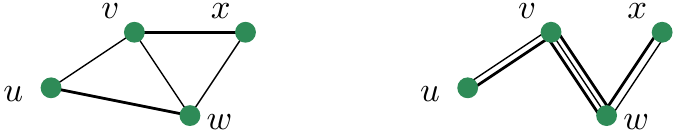}
\caption{Path endpoint ambiguity.  A connection exists between $(u, w)$ and $(v, x)$ and there is a path between $u$ and $x$. When edges are bundled along paths, the viewer could perceive a direct edge between $(u, x)$ if the bundling is strong.  %
}
\label{fig:pathamb}
\end{figure}

{\bf Path Endpoint Ambiguity.}  Edge-Path bundling does not create connections between vertices that do not exist in the underlying graph.  However, it can produce ambiguities in edge endpoints along a path.  Fig.~\ref{fig:pathamb} shows this ambiguity.   A path exists, by definition between $u$ and $x$, but it can be difficult to tell if there is a direct connection between $u$ and $x$ in the Edge-Path bundle, depending on bundling strength.

{\bf Edge Crossing Ambiguity.}  When two edges or bundles cross, there could be an ambiguity if the crossing angle is shallow~\cite{hhe-eca-08}.  This is a fundamental ambiguity of graph drawing in general, which also affects the unbundled straight-line input drawing $D_G$.

\section{Quality Metrics}
In order to evaluate the edge bundling created by different bundling algorithms, we make use of three quantitative metrics. They measure the amount of clutter reduction using the \emph{ink reduction}, the \emph{distortion} of edge lengths in the graph, and the amount of adjacency \emph{ambiguity} in the bundled layout.  This section defines these quality metrics. We remark that no single quality metric can fully judge a bundled layout, but we think that a good bundled layout, which reduces visual clutter but at the same time aims to be faithful to distances and graph topology, will do fairly well on all three metrics. 

\subsection{Ink Reduction}

Let $I$ be the grayscale bitmap image $I \in \{0,\dots,255\}^{m \times n}$ of a bundled graph layout $\Gamma$ and let $I^B \in \{0,1\}^{m \times n}$ be the binarization of $I$ such that

\begin{equation}
    I^B(i,j) = \begin{cases}
                        1 & I(i,j) \ge \delta\\
                        0 & I(i,j) < \delta
                \end{cases},
\end{equation}
where $0 \le \delta \le 255$ is a global gray value threshold to consider a pixel occupied. Similarly we define $J$ as the unbundled input grayscale image.  The ink-reduction of $I$ with respect to $J$ is defined as:

\begin{equation}
    \mathrm{ink}_J(I) = \frac{\sum_{i=1}^m \sum_{j=1}^n I^B(i,j)}{\sum_{i=1}^m \sum_{j=1}^n J^B(i,j)}. 
\end{equation}

Under the assumption that the bundling reduces the number of pixels occupied by ink, the ink reduction takes a value between $0$ and $1$ and measures the factor by which the number of pixels occupied in the graph layout is reduced. An ink reduction close to $1$ indicates a low degree of bundling, whereas smaller values indicate a higher degree of bundling. It is possible to obtain values larger than $1$ if the bundling increases the number of pixels occupied.

\subsection{Distortion}

Let $e = (u, v) \in E$ be an edge in $G$ with Euclidean length of $||u - v||$ and let $d_\Gamma(u,v)$ be the length of the curve connecting $u$ and $v$ in the bundled layout $\Gamma$. We define the distortion of layout $\Gamma$ as the average distortion of its edges: %
\begin{equation}\mathrm{dist}(\Gamma) = \frac{1}{|E|} \sum_{(u, v) \in E} \frac{d_\Gamma(u,v)}{||u-v||} %
\end{equation}

The distortion measures the factor by which 
\hedit{length} of an edge increases in the bundled layout on average. Distortion values 
\hedit{near}
$1$ mean that 
\hedit{lengths of bundled edges remain}
close to the Euclidean distance of their 
\hedit{endpoints; larger values mean}
edge bundles have longer detours from straight line distance, making adjacencies harder to read~\cite{heh-grbgt-09}.

\subsection{Ambiguity}

\begin{figure}
\centering
\includegraphics[width=0.9\linewidth]{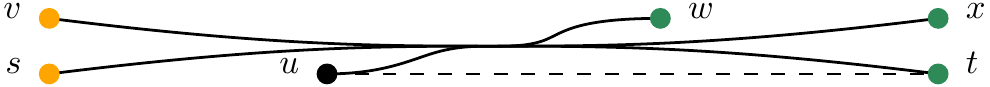}
\caption{Consider an edge $e$ with endpoints $(s,t)$. The bundling implies $N_{\Gamma}(s,e)=\{t, w, x\}$ and $N_{\Gamma}(t,e)=\{s, u, v\}$. Additionally, there exists an edge with endpoints $(u, t)$ and therefore we would consider $u$ a true neighbour of $t$ resulting in $N^f_{\Gamma}(s,e)=\{w, x\}$ and $N^f_{\Gamma}(t,e)=\{v\}$}
\label{fig:pb1}
\end{figure}

We define %
an ambiguity metric, inspired by faithfulness~\cite{neh_ofgv_2013}, to measure the number \hedit{and severity of perceivable false connections in a bundled layout (Fig.~\ref{fig:pb1})}.  Let $e = (s,t) \in E$ in a bundled layout $\Gamma$. We define the set of reachable neighbors of the endpoint $s$ along $e$ as $N_{\Gamma}(s,e) = \{v \in V \mid \exists \text{ ambiguous connection from } s \text{ to } v \text{ in } \Gamma\}$; analogously we define $N_{\Gamma}(t,e)$ for the endpoint $t$. \hedit{An ambiguous connection occurs if another edge $e'=(u,v)$ intersects or is closer than a distance threshold $\epsilon$ at a point $p$ on $e$ in $\Gamma$ and the angle at $p$ between $e$ and $e'$ is smaller than a threshold $\theta$.} %
\hedit{Intuitively, edges $e$ and $e'$, locally in $p$, are difficult to distinguish and thus the endpoint of $e'$, say $v$, forming the small angle with $t$ is a reachable neighbor in $N_{\Gamma}(s,e)$, whereas the other endpoint $u$ belongs to $N_{\Gamma}(t,e)$, independently of $(s,v)$ or $(u,t) \in E$ or not.}

The reachable neighbor sets contain some true and false neighbors and we may take this classification based on a \hedit{graph distance threshold} $\delta \ge 1$: the set of true neighbors is defined as $N_\Gamma^t(s,e) = \{v \in N_\Gamma(s,e) \mid d_G(s,v) \le \delta\}$ and the set of false neighbors as $N_\Gamma^f(s,e) = N_\Gamma(s,e) \setminus N_\Gamma^t(s,e)$. 
\hedit{Here let $d_G(s,v)$ denote the hop distance between $s$ and $v$ in $G$, i.e., the length of the shortest unweighted path between $s$ and $v$ in $G$.
}
We can now define the ambiguity of $\Gamma$:
\begin{equation}\label{eq:edgeamb}\mathrm{amb}(\Gamma) = \frac{\sum_{v \in V} \sum_{e=(v,w) \in E} |N_\Gamma^f(v,e)|}{\sum_{v \in V} \sum_{e=(v,w) \in E} |N_{\Gamma}(v,e)|}.\end{equation} 
\hedit{
This value measures the proportion of false neighbors to all neighbors implied by $\Gamma$, with low values corresponding to less ambigous drawings.}
\hedit{
For $\delta=1$ %
the set $N_\Gamma^t$ contains only the actual neighbors in edge set $E$, while $\delta\rightarrow\infty$ counts all reachable vertices as true neighbors; the remaining false neighbors correspond to vertex pairs in different connected components (i.e. the most ambiguous). 
Hence for connected graphs, there is some value $\delta_{0} \in \mathbb N$ for which $\mathrm{amb}(\Gamma)$ drops to zero for all $\delta \ge \delta_{0}$.
Values for $\delta > 1$ %
correspond to tolerating ambiguous connections if the endpoints are connected by paths of at most $\delta$ edges. We use $\mathrm{amb}^\delta(\Gamma)$ to denote the ambiguity for a particular value $\delta$.}

This ambiguity measure will count all ambiguities due to shallow edge crossings, independent edge ambiguities (Fig.~\ref{fig:ambcase}), and path endpoint ambiguities (Fig.~\ref{fig:pathamb}).  Thus, Edge-Path bundling will have a non-zero value for this measure.  However, in all of the Edge-Path bundling drawings, by definition, there are no independent edge ambiguities.

\subsection{Detecting Edge Ambiguity in a Drawing}\label{sec:detectambiguity}
The approach chosen to detect edge ambiguity is based on the premise that nearby edges with a similar direction may mistakenly have their endpoints interchanged, \hedit{especially edges} that are parallel or cross with a small angle. 
\hedit{We thus check}
if edges are spatially close and if the crossing angle is below the threshold $\theta = 7.5^\circ$, a crossing angle for which empirical evidence~\cite{hhe-eca-08} shows a strong negative effect on readability. 

First, the drawing area is divided into a small, square grid. Two edges are considered ambiguous if an ambiguity is detected in at least one cell. Each grid cell is assigned the intersecting edge segments it contains as well as their respective angle. Internally, curved edges are approximated using polylines, leading to two cases. In the first case, a segment ends in or intersects a grid cell and the angle of the segment is assigned. In the second case, where multiple segments of the polyline intersect the grid cell, the mean of the angles is assigned, under the assumption that the grid cell is small enough such that there is no drastic change of direction.

After this initial assignment, a sliding window processes the cells of the grid. The angles of the assigned segments of an edge are aggregated over all cells in the window which gives us an angle for each edge intersecting the window. Edges are processed in a pairwise manner. Given that two edges intersect the window, the edges must be spatially close. Therefore, the smaller of the two crossing angles is compared to a threshold $\theta$ to determine if it is an ambiguity. 

Additionally, we need the set of neighbours for the endpoints of an edge $e=(s,t)$ in the sets $N_\Gamma(s,e)$ and $N_\Gamma(t,e)$. When assigning segments to grid cells we iterate over the segments of an edge from source to target and assign an angle in the range $[0,2\pi]$. This gives enough information to determine the relative position of ambiguous edge segments by comparing their induced angles and deciding if the source vertex of one edge is ambiguous with the source or target vertex of another edge. This is analogously defined for the target vertex.

\section{Data} \label{sec:data}

We computed bundled layouts of real and synthetic datasets. The synthetic datasets highlight properties of bundling behaviour. The real world datasets are common in the literature and used for comparison.

\begin{figure}[htb]
    \centering
    \includegraphics{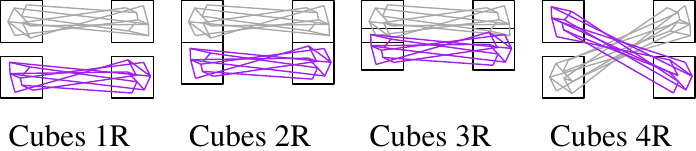}
    \caption{Illustration of the Cubes 1R--4R datasets.}
    \label{fig:cubes_data}
\end{figure}

{\bf Cubes 1R-4R.} We created several synthetic datasets that should capture connections between disconnected components, a common feature in real-world datasets. Vertices ($|V| = 100$) are evenly distributed into four components which are subsequently randomly embedded inside axis-parallel cubes with fixed side length $s$. The cubes are positioned top left, top right, bottom left and bottom right with fixed space of $2s$ between the right edges of the left cubes and the left edges of the right cubes, see Fig.~\ref{fig:cubes_data}. We introduce a distance $\Delta$ between the bottom edge of the top cubes and the top edge of the bottom cubes. In Cubes 1R and 4R $\Delta=s/10$, which creates space between the cube pairings. In Cubes 2R $\Delta=0$ and for Cubes 3R $\Delta=-s/5$ which results in overlapping top and bottom cubes. A random spanning tree is used to connect vertices inside their cube component and additional edges are randomly added. In Cubes 1R-3R the left cubes are connected with their right counterpart and in Cubes 4R the components are connected diagonally. For connecting the components we added $|V|/2$ edges randomly. In the directed variation we additionally specify half of the edges going from left to right and the other half from right to left.%

{\bf Noise.} This dataset is based on the observation by Lhuillier et. al~\cite{lhuillier_2017} that bundling approaches do bundle without support in the underlying data. For the Noise dataset, we embedded $|V| = 1000$ vertices randomly in  a square and connected vertices to form a perfect matching, resulting in $|E| = |V|/2$ edges and connected components. 

{\bf US Airlines.} The US Airlines dataset has been introduced by~\cite{holten_2009} and is commonly used in bundling publications. The dataset depicts flight paths from source to target airports in the US and has $|V| = 235$ vertices and $|E| = 2101$ edges. \hedit{It has one connected component.}

{\bf Migrations.} This dataset~\cite{cui_2008} shows migration in the US~\cite{migrations} and consists of a set of trails which we converted into a graph. The directed graph has $|V| = 1702$ vertices and $|E| = 9726$ edges. The number of vertices is slightly lower compared to earlier literature (1712) when we converted the trail sets. The undirected graph has $|E| = 6487$ edges. \hedit{It has 28 connected components.}

{\bf Air Traffic.} This dataset consists of global flights and has $|V| =1533$ vertices and $|E| = 14825$ edges. \hedit{It has one connected component.}

\hedit{{\bf Amazon Subset.} This dataset consists of products with edges indicating that they are commonly co-purchased. We randomly filtered edges from the original graph~\cite{snapnets,LeskovecAH06}. It has $|V| = 192k$ vertices and $|E| = 269k$ edges.}

\section{Experiment Rendering and Bundling Algorithms}

To guarantee a fair comparison between the different systems, we rendered all images of the different bundling results in the same way. First, we computed a bundling with the systems' respective implementation. Then, we extracted the bundled edges as polyline approximations of the curves and used this as input for our own rendering. As some systems required a different scaling of the input layout, we proceeded to scale all bundlings to a fixed width of $1600$px while keeping the original aspect ratio.
In the standard bundling plot, we used a linewidth of $1$px to draw the edges and a diameter of $4$px for disks representing vertices. 
We used this plot to calculate the ink reduction of a bundling approach. 

To give a better impression of the direction of edges in a bundle, we created a second plot where we applied a perceptually uniform colour map~\cite{crameri_2020} to assign a color to each edge depending on 
\hedit{its angle}
in the straight line drawing. Furthermore, we created plots of the distortion, where we first computed the minimum and maximum distortion over all results of a dataset and applied a sequential colormap~\cite{crameri_2020} to show an overview of where edges are distorted in a bundling. Finally, as we calculate the ambiguity of edges on a per cell basis, we can convert this to an image. We assign each cell the number of intersecting ambiguous edge-pairs and normalize by the maximum over all results of a dataset to achieve a greyscale image that encodes areas of high ambiguity as white spots. These images can be found in the supplementary material. %

\subsection{Bundling Algorithms in the Experiment}
For our experiment, we selected  algorithms with available implementations as representatives from major categories of bundling approaches: force-directed bundling, confluent drawings with fixed vertex positions, grid-based approaches, and image-based approaches. 

\textbf{Force-directed Bundling.} We used an implementation in D3.js~\cite{upphiminn2021Aug}
and set the parameters as specified in~\cite{holten_2009} (spring constant $K =0.1$, force iterations $I=60$, subdivision operation $C=6$, initial subdivision points $P=1$, increase ratio $2$). As the bundling depends on the scale of the input graph, we did not rescale the input layout before bundling.

For \textbf{Confluent Drawings.} we used an available implementation~\cite{jxz122021Aug} 
based on the results of~\cite{bach_2016, zheng_2021}, which we modified to handle graphs with an embedding. This was realized by removing the layout step of the implementation and replacing it with the following. First, we fixed all vertices of the input layout and embedded the routing vertices in the barycenter of neighbouring input vertices. Afterwards, we iteratively moved the routing vertices towards the the barycenter of their neighbours until the layout converged.  Generally, confluent bundling algorithms draw and bundle the graph simultaneously, but our approach requires a drawing as input.  This means that these approaches are hampered as they cannot modify the layout.

\textbf{Winding Roads.} We used the implementation~\cite{lambert_2010} in Tulip~\cite{auber_2018}. 

\textbf{KDEEB.} We adapted an available implementation~\cite{Hurter2019Jul}
of KDEEB~\cite{hurter_2012}. A comparison with FFTEB~\cite{lhuiller_2017} would be preferable, but we were not able to run the implementation. Furthermore, given the computational effort required for our larger experiments we used the GPU implementation which had its parameters fixed by the authors. The layout of the input graph was scaled to the range $[0,1]$.

\hedit{\textbf{CUBu} We used the available implementation~\cite{Telea2021Cubu} 
of CUBu~\cite{vdzwan_2016}. A medium sized kernel was used while most default settings where kept. We increased relaxation to get smoother lines.}

\textbf{Divided Edge Bundling.} We used an implementation in Matlab~\cite{kakearney2021Aug} 
of Selassie {\it et al.}~\cite{selassie_2011} for directed edge bundling using default parameters. %

\subsection {Parameter Specification}

As discussed in Sec.~\ref{sec:algorithm}, we can specify several parameters when computing a bundling. We tested different settings and evaluated the resulting ink reduction, distortion, and ambiguity. 

We found that a maximum distortion factor $k \in [1.5, 3.0]$ works well for most of the experiments. A distortion factor of less than 1.5 resulted in very little edge bundling. For the \hedit {Cubes datasets}, a distortion factor above 3.0 resulted in overaggregation with ink reduction $> 1$. %
In the US Airlines dataset, when increasing the distortion factor more edges were bundled, but increasing above a factor of 3.0 did not produce more bundling, which can be explained by the fact that most shortest paths are below the distortion threshold already. Images of the effect of this parameter on Airlines are contained in the supplementary material. Ultimately, we used $k = 2.0$ for all experiments.

Experiments on the edge weight factor $d$ showed that values of $d \in \{1,2,3\}$ work well. Higher values of $d$ penalise long edges from being included in shortest paths used for bundling.  A value of $d=2$ was chosen for our experiments.

As our algorithm computes a list of control points for each edge, we can use an integer smoothing parameter to pull the rendered curve towards its control points. A smoothing parameter of 1 would use the original list of control points to calculate the curve. A smoothing parameter of $2$ would add an additional control point in the middle of the line between every consecutive pair of control points. \hedit{Increasing the smoothing factor above $2$ results in recursively adding more control points thus creating a tighter bundling.}  In our experiments, we set the smoothing parameter to $2$. 

\section{Experiments}

We next
discuss our experimental results.  KDEEB and CUBu are 
image-based techniques.  As it was easier to adjust the kernel size of CUBu, its result images are 
shown.
High resolution images of all 
algorithms, including KDEEB, are in the supplementary material.

\subsection{Runtimes}
\hedit{We used all the cited implementations specified above.  As these implementations are written in a variety of languages with some designed to run on the GPU and others the CPU, direct comparison of runtimes is less informative.  However, it was important to note that the image-based techniques, KDEEB and CUBu, have by far the best performance in terms of runtime, processing large graphs in less than a second.}

\hedit{We report the average runtimes of ten executions for Edge-Path bundling.  We implemented our bundling algorithm in C++ and ran it on a machine with Ubuntu 20.04 operating system, AMD Ryzen 5 5600x CPU and NVidia RTX 3070: Cubes (1-4)R [128ms], Noise [401ms], Airlines [920ms], Migrations [undirected: 10.6s, directed: 15.3s], Air Traffic [49.6s], and Amazon Subset [7.5 hours].}

\hedit{Furthermore, we also report runtimes for bundling the Air Traffic dataset, the largest dataset presented in an image, with CUBu [5ms incl. rendering], KDEEB [45ms], Force-directed [32.5s], Force-directed (divided) [3.2hrs], Winding Roads [7.6s], and Confluent [157s].}

\subsection {Synthetic Data}

\begin{figure*}
    \centering
      \subfigure[Straight Line]{\label{fig:cubesSL}\includegraphics[width=0.16\linewidth]{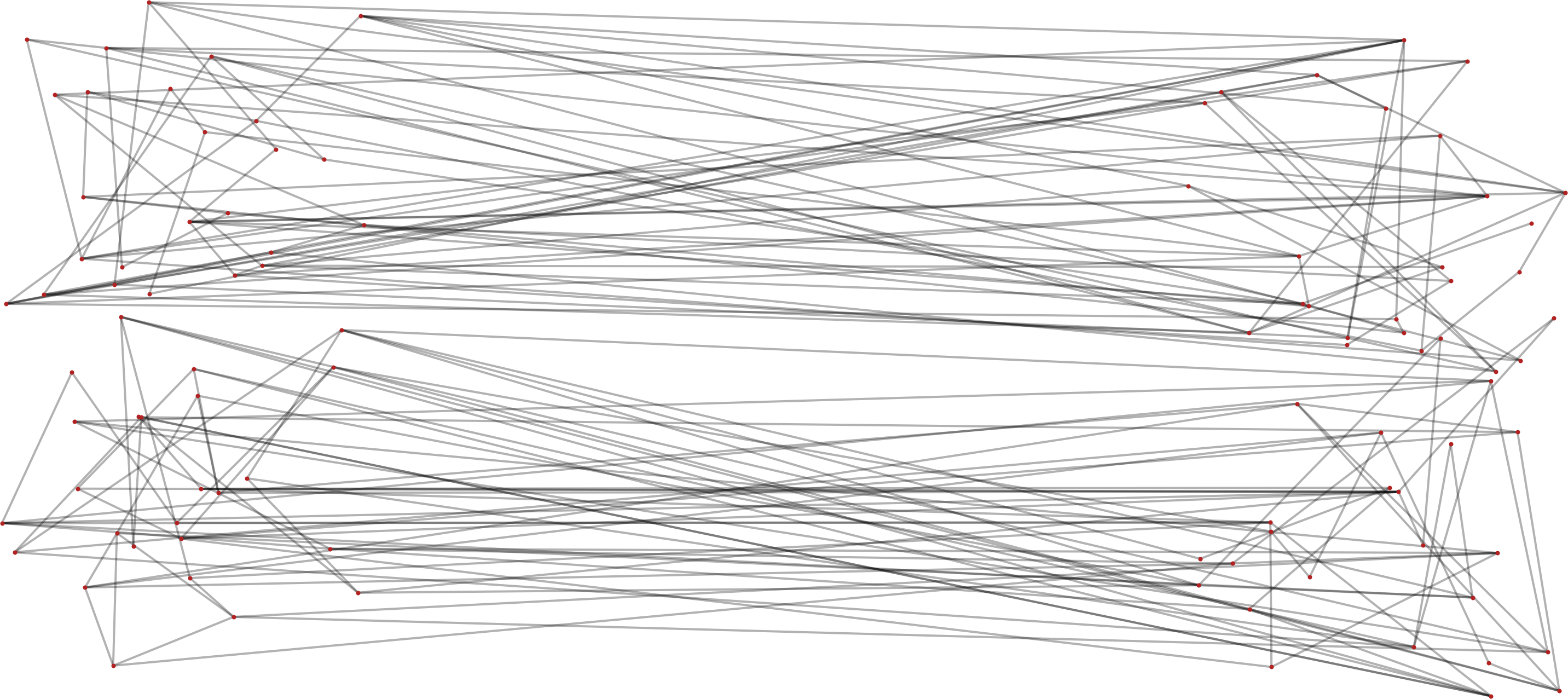}}
      \subfigure[CUBu]{\label{fig:cubesCUBu}\includegraphics[width=0.16\linewidth]{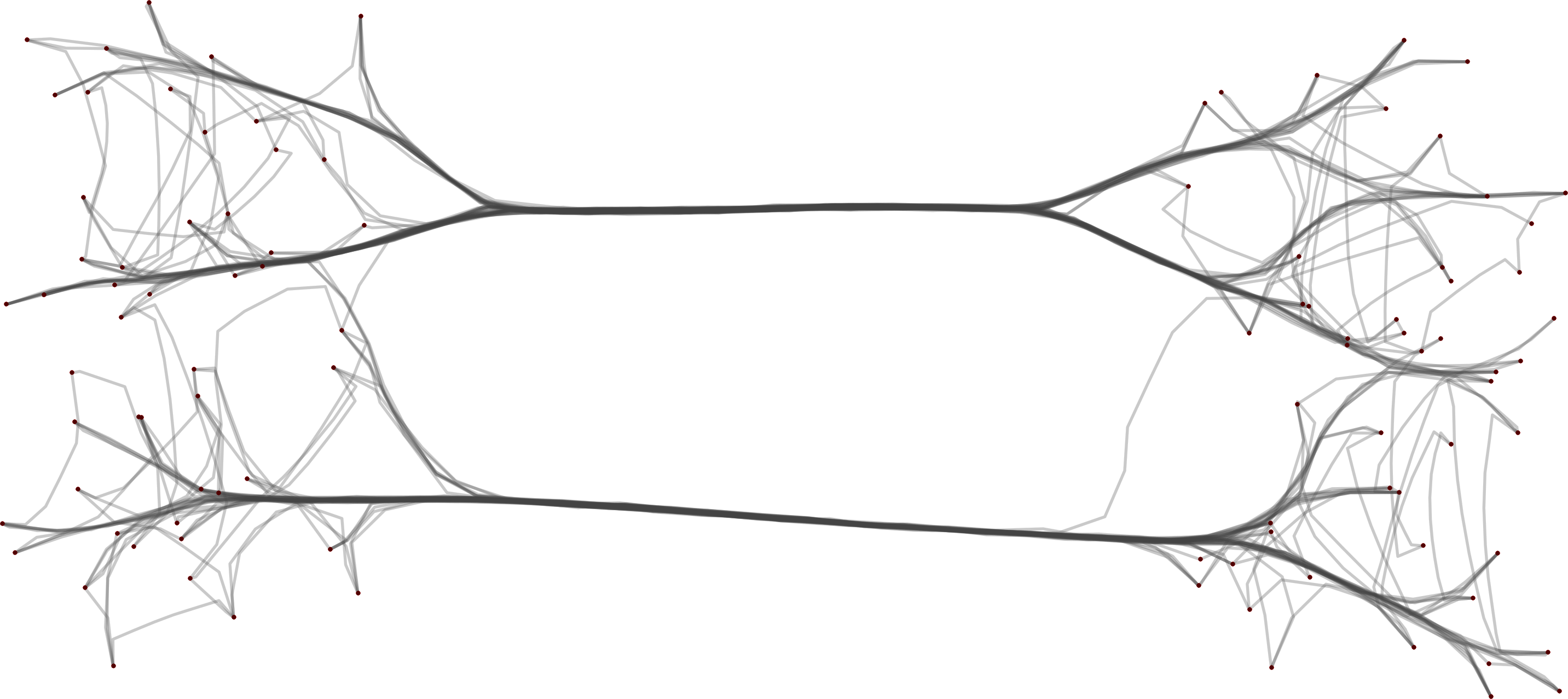}}
      \subfigure[Winding Roads]{\label{fig:cubesWR}\includegraphics[width=0.16\linewidth]{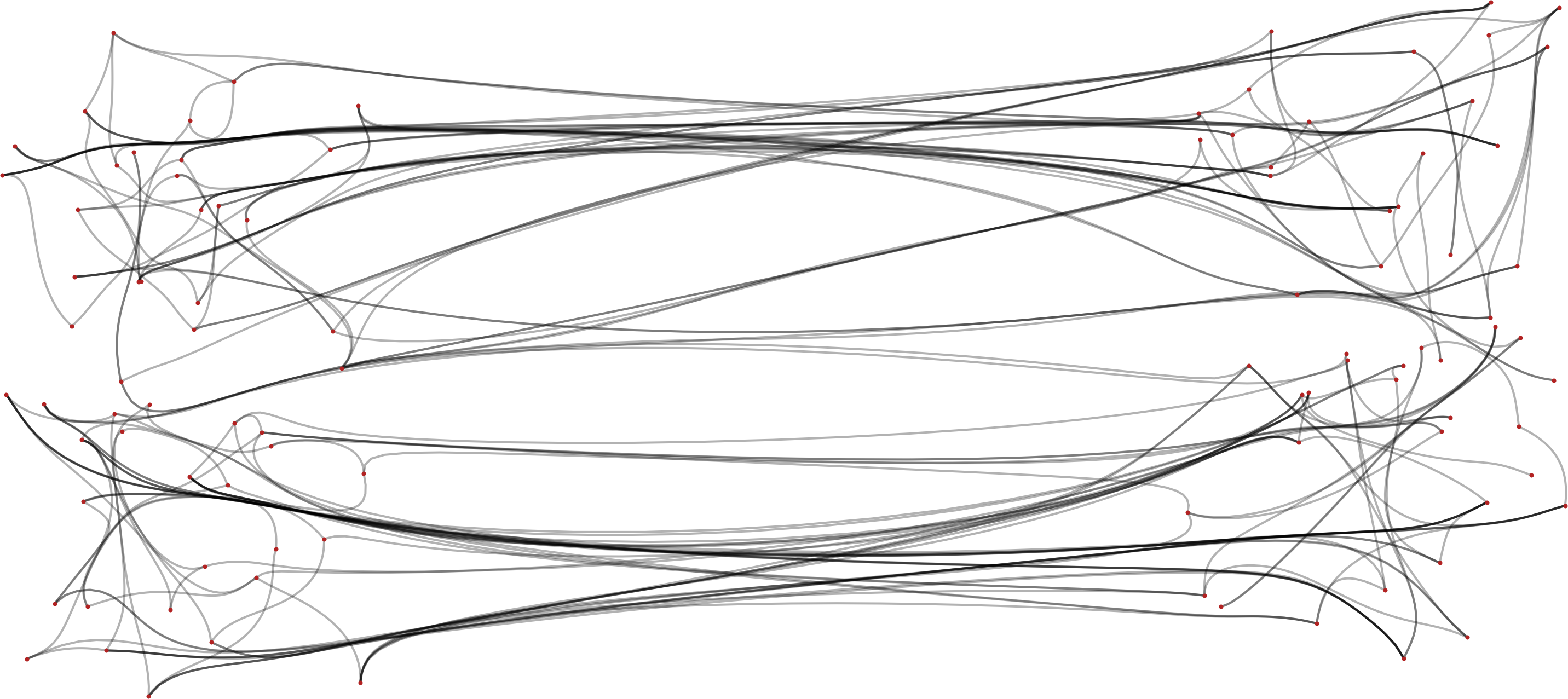}}
      \subfigure[Confluent]{\label{fig:cubesConf}\includegraphics[width=0.16\linewidth]{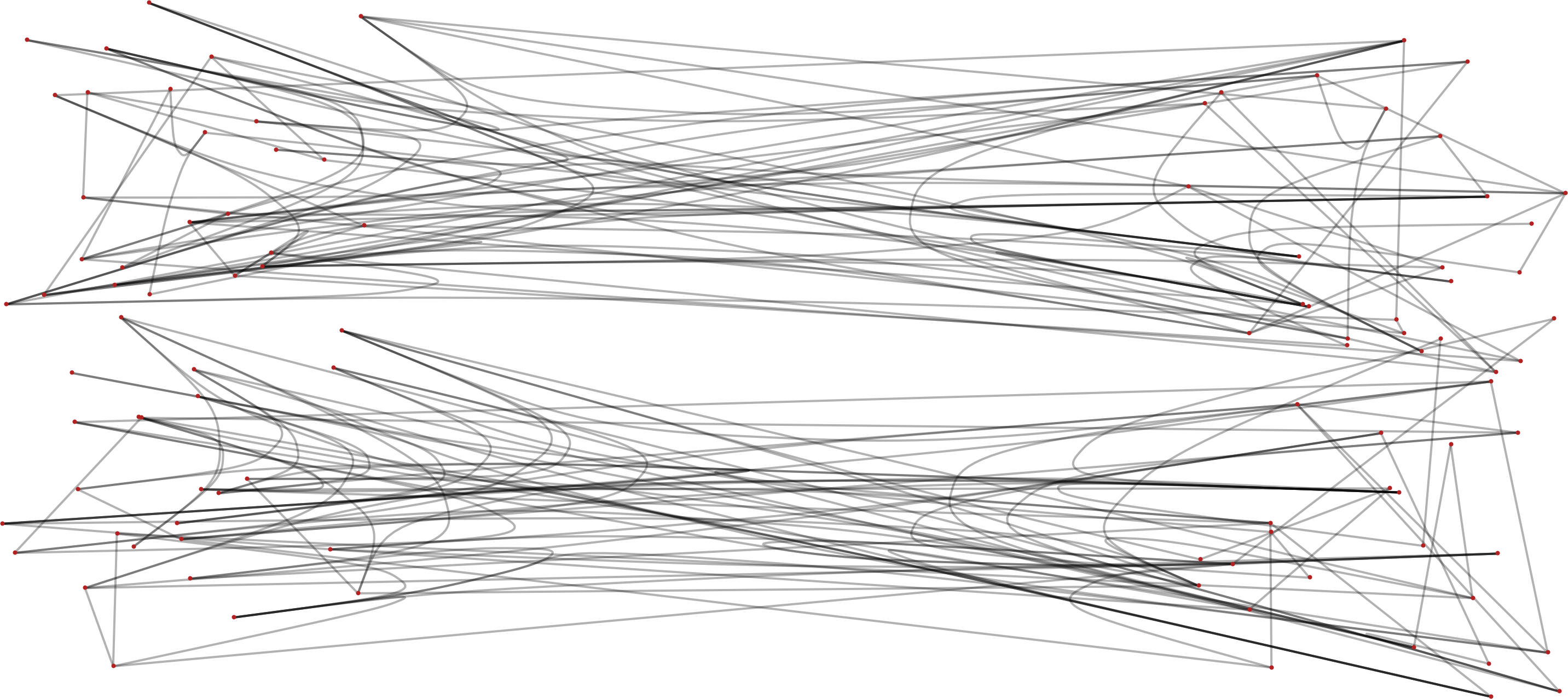}}
      \subfigure[Force-Directed]{\label{fig:cubesFB}\includegraphics[width=0.16\linewidth]{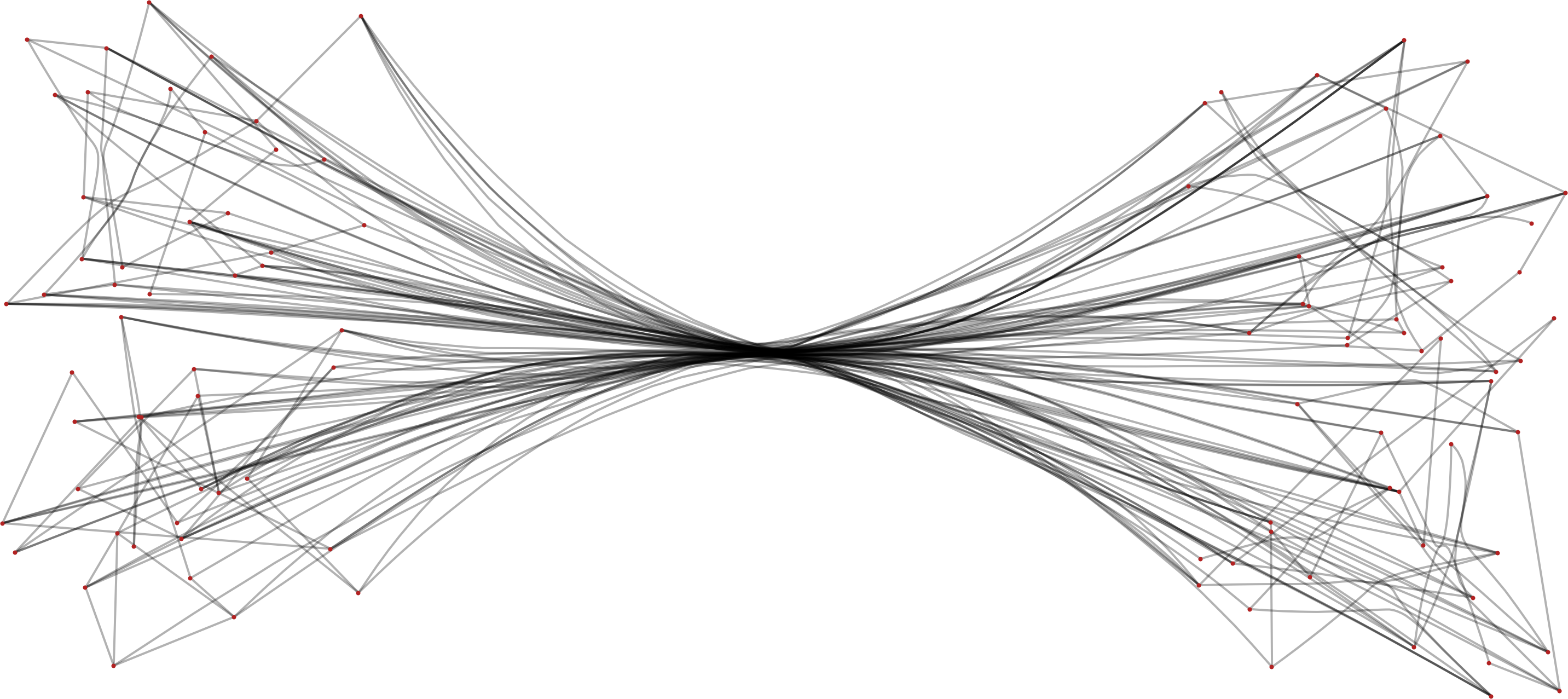}}
      \subfigure[Edge-Path]{\label{fig:cubesFPB}\includegraphics[width=0.16\linewidth]{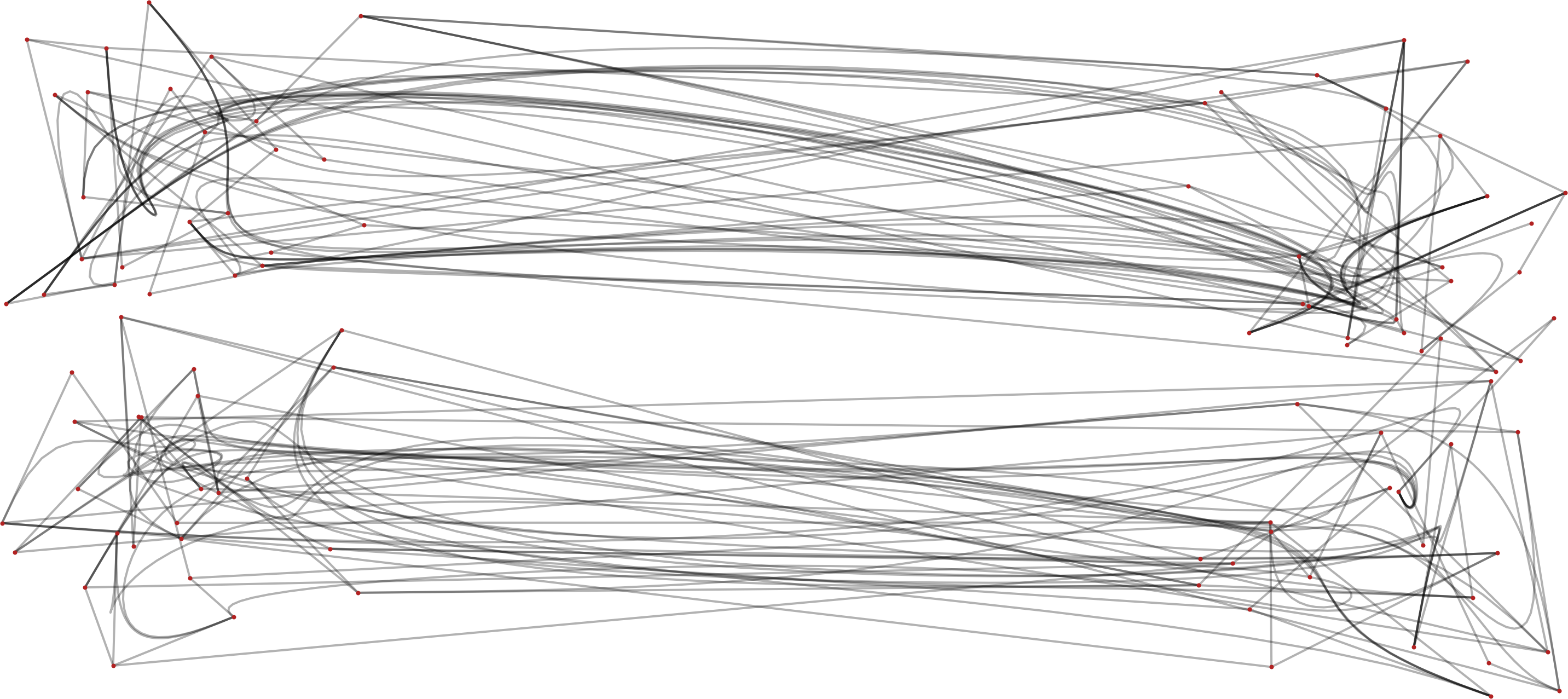}}
    \caption{The Cubes 2R dataset shows the bundling of two connected components. KDEEB, Winding Roads and Force-Directed bundle edges regardless if edges are connected in the data. Confluent drawings and Edge-Path only bundle edges in their respective components.}
    \label{fig:cubes2R}
\end{figure*}

We first discuss the quantitative results on our synthetic data, Cubes as well as the Noise data. By design of the Cubes datasets, there are two disjoint edge sets from left to right, which can potentially be grouped into distinct bundles, but edges from disconnected components should ideally not be mixed.  In the Noise data, consisting of disconnected edges, there is nothing to be bundled, topologically speaking. Refer to Table~\ref{tab:synthetic} for the scores of the quality metrics and Fig.~\ref{fig:cubes2R} for the result images.  \hedit{The results for Cubes 1R and 2R were similar to 3R and 4R.  Full metric results available in the supplementary material.}

\begin{table*}[t]
\small
    \caption{Scores of the quality metrics for the undirected synthetic datasets and all bundling algorithms. Column  $\mathrm{dist}_\Gamma$ gives mean and median. Columns $\mathrm{amb}^\delta$ are only shown for $1 \le \delta \le 5$ if there are non-zero entries. For Noise, all $\mathrm{amb}^\delta$ are equal as the graph consists of independent edges.
    }
    \centering
    \setlength{\tabcolsep}{1.2ex}
    \begin{tabular}{r||c|cc|c|c|c|c||c|cc|c|c|c|c|c||c|cc|c}
                            & \multicolumn{7}{c||}{Cubes 3R}   & \multicolumn{8}{c||}{Cubes 4R}  & \multicolumn{4}{c}{Noise} \\
                            &  $\mathrm{ink}_J$   &  \multicolumn{2}{c|}{$\mathrm{dist}$}     &  $\mathrm{amb}^1$ &  $\mathrm{amb}^2$ &  $\mathrm{amb}^3$ &  $\mathrm{amb}^4$ 
                            &$\mathrm{ink}_J$   &  \multicolumn{2}{c|}{$\mathrm{dist}$}   &  $\mathrm{amb}^1$    & $\mathrm{amb}^2$ &  $\mathrm{amb}^3$ &  $\mathrm{amb}^4$ &  $\mathrm{amb}^5$  & 
                            $\mathrm{ink}_J$    &  \multicolumn{2}{c|}{$\mathrm{dist}$}     &  $\mathrm{amb}^{\delta}$ \\
                            \hline
        Straight-Line       & 1.00 &  1.00& 1.00 &  0.50 & 0.32 & 0.08 & 0.01                   & 1.00 &  1.00& 1.00 &  0.56 & 0.37 & 0.12 & 0.02 & 0.00               & 1.00 &  1.00& 1.00 &  \textbf{0.50} \\
        \hline
        Force      & 0.79 & \textbf{1.04}& 1.02 & 0.87  & 0.61 & 0.19 & 0.01                             & 0.70 & \textbf{1.01}& \textbf{1.00}  & 0.88 & 0.65 & 0.30 & 0.12 & 0.10        & 0.98 & \textbf{1.00}& \textbf{1.00} & 0.67   \\
        Confluent           & 0.93 & 1.23& \textbf{1.00} & \textbf{0.73} & \textbf{0.42} & \textbf{0.13} & 0.01            & 0.93 & 1.20& \textbf{1.00}  & \textbf{0.79}  & 0.54 & 0.28 & 0.16 & 0.14               & 1.00 &  \textbf{1.00}& \textbf{1.00} &  \textbf{0.50}  \\
        WindingR       &  0.50 & \textbf{1.04}& 1.03 & 0.81 & 0.55 & 0.19 & 0.01                         & 0.51 & 1.04& 1.02 & 0.81 & 0.57 & 0.25 & 0.09 & 0.05               & 0.64 & 1.06& 1.05 & 0.96   \\
        KDEEB               & \textbf{0.22} & 1.11& 1.07 & 0.90 & 0.64 & 0.22 & 0.01          & \textbf{0.26} & 1.13& 1.09 & 0.94 & 0.81 & 0.60 & 0.48 & 0.46        & \textbf{0.43} & 1.19& 1.18 & 0.99      \\
        CUBu & 0.28 & 1.08 & 1.07 & 0.90 & 0.63 & 0.20 & 0.01                                   & 0.32 & 1.10 & 1.09 & 0.95 & 0.81 & 0.60 & 0.49 & 0.47 & 0.54 & 1.16 & 1.15 & 0.99 \\
        Edge-Path           & 0.84 & 1.05& \textbf{1.00} & 0.80 & 0.49 & \textbf{0.13} & 0.01            & 0.91 & 1.04& \textbf{1.00} & \textbf{0.79}  & \textbf{0.50} & \textbf{0.16} & \textbf{0.02} & \textbf{0.00}      &  1.00 &  \textbf{1.00} & \textbf{1.00} &  \textbf{0.50} \\
        
    \end{tabular}
    \label{tab:synthetic}
\end{table*}

{\bf Ink Reduction.} Naturally, the ink reduction for Straight Line is~1. KDEEB and CUBu consistently and as expected reduce 
\hedit{ink}
the most on all four Cubes instances. Winding Roads is third and then Force-Directed. The stronger ink reduction of the other approaches comes at the cost of overbundling some edges and creating ambiguities. Confluent and Edge-Path have a relatively low 
\hedit{reduction; this}
is expected as the more \hedit{graph topology-aware bundling algorithms do not create independent edge ambiguities.}  
\hedit{Results}
on Noise are similar (see also Fig.~\ref{fig:noise}), where KDEEB and CUBu, Winding Roads, and then Force-Directed reduce 
\hedit{ink}
the most. Confluent and Edge-Path do not change the input layout at all, as they never bundle independent edges.

{\bf Distortion.} Straight Line, by definition, has a distortion of 1. We report mean and median distortion values. While the mean distortion is generally least for Force-Directed bundling with just a few percent, the median distortion is consistently smallest for Edge-Path, meaning that the majority of edges are unbundled and thus undistorted. Winding Roads is the next best, with CUBu, and KDEEB. Although Confluent has a high mean distortion, its median is 1 indicating it distorts a few edges within the components drastically as the layout cannot be adjusted.
The observed distortions in the Noise data is non-existent for Confluent and Edge-Path as there is no bundling. Force-Directed, producing only few and thin bundles, has generally low distortion, followed by Winding Roads, CUBu, and KDEEB. 

{\bf Ambiguity} The ambiguity scores are lowest for Straight Line as it has no bundling with values of 0.47--0.56 caused by shallow-angle edge crossings only.  
 All bundling algorithms increase the ambiguity with KDEEB, CUBu, and Force-Directed having the largest ambiguity scores followed by Winding Roads and Edge-Path. Confluent has the lowest ambiguity score.  Algorithms that produce stronger bundling with more ink reduction often produce more independent edge ambiguities. With increasing $\delta$, Edge-Path becomes the best performing algorithm in terms of ambiguity. %
As expected, on the Noise data, Confluent and Edge-Path again obtain the same initial ambiguity as Straight Line. Force-Directed has the next lowest ambiguity score followed by Winding Roads and KDEEB.

{\bf Qualitative.}  Fig.~\ref{fig:cubes2R} shows our results.  On the Cubes datasets, Force-Directed, KDEEB, CUBu and Winding Roads can mix the two independent streams of edges. Confluent does not mix these streams by design, but has a low degree of bundling.  Edge-Path bundling bundles within each stream but does not connect the two disconnected streams.  This demonstrates that it is able to avoid grouping unrelated edges together.

\subsection {Real Data}
\begin{figure*}[p]
    \centering
      \subfigure[Edge-Path]{\label{fig:migrationComparisonEBP}\includegraphics[width=0.3\linewidth]{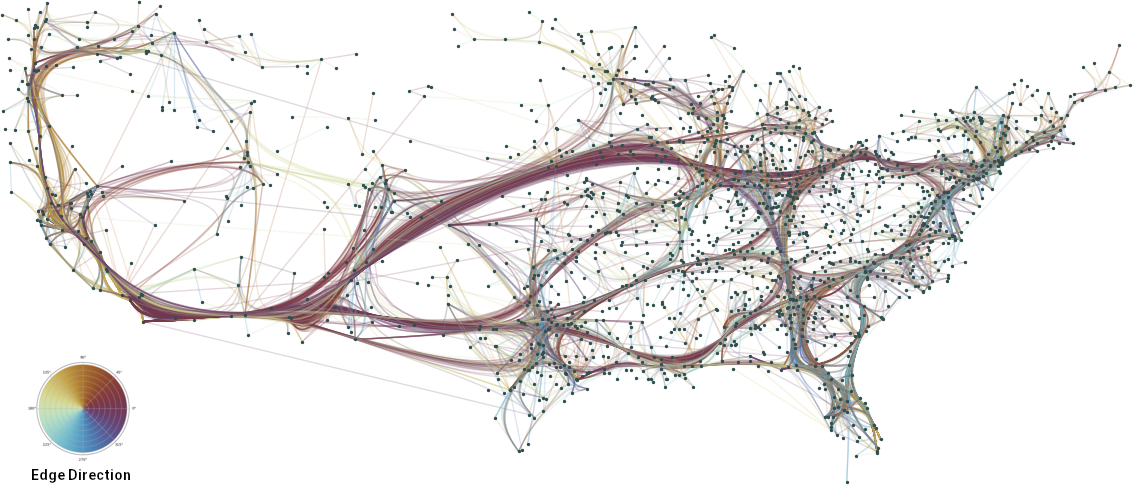}}
     \subfigure[Force-Directed (Divided)]{\label{fig:migrationComparisonFBD}\includegraphics[width=0.3\linewidth]{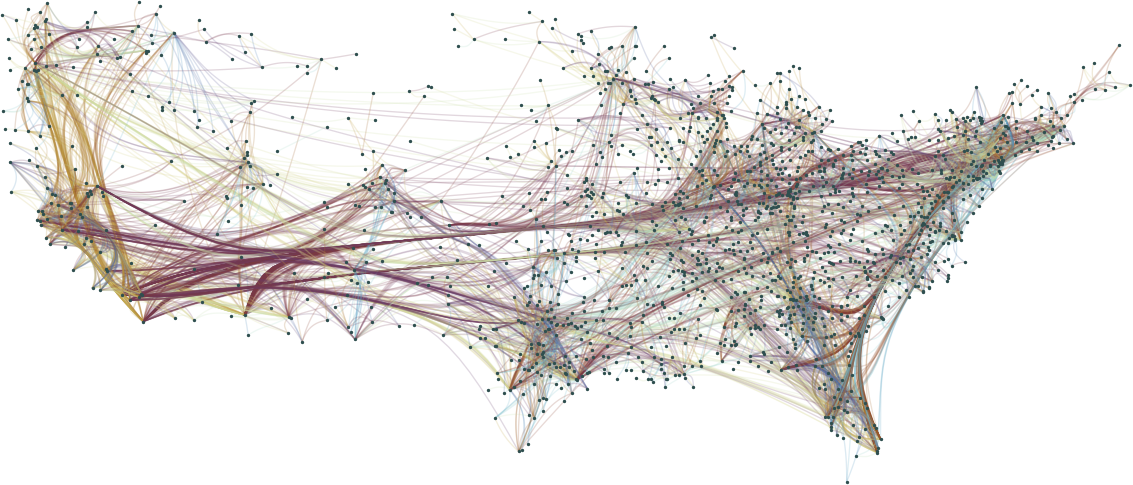}}
      \subfigure[CUBu]{\label{fig:migrationComparisonCUBu}\includegraphics[width=0.3\linewidth]{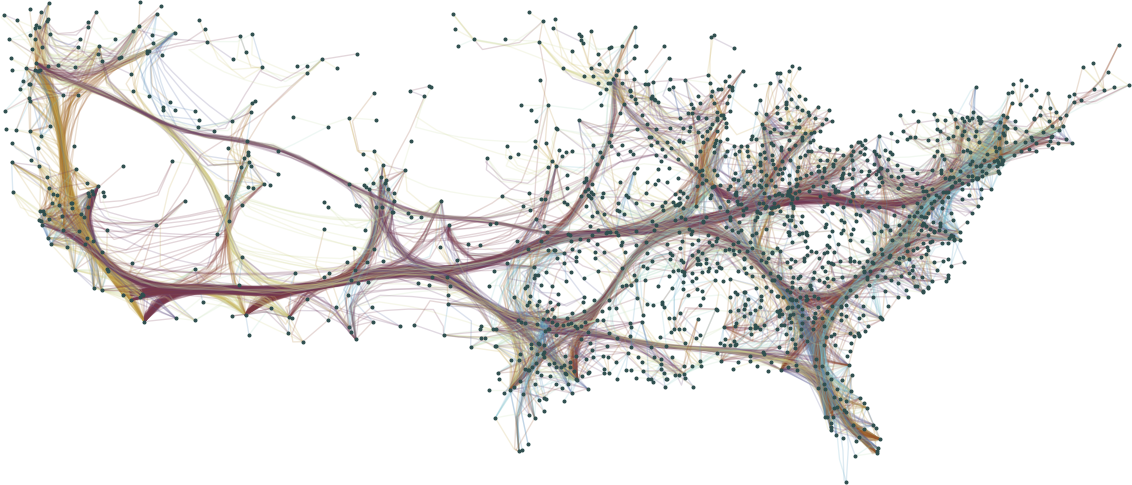}}
    \caption{Migrations (directed). {{\bf (a)} Directed Edge-Path bundling. \bf (b)} Divided edge-bundling which uses a forced-based approach. {\bf (c)} CUBu.  Directed Edge-Path bundling does not bundle unrelated edges together and can reveal more detail on the east coast and in the east-west flow.}
    \label{fig:migrationsDirected}
\end{figure*}

\begin{figure*}[p]
    \centering
        \subfigure[Force-Directed]{\label{fig:airtrafficFBD}\includegraphics[width=0.47\linewidth]{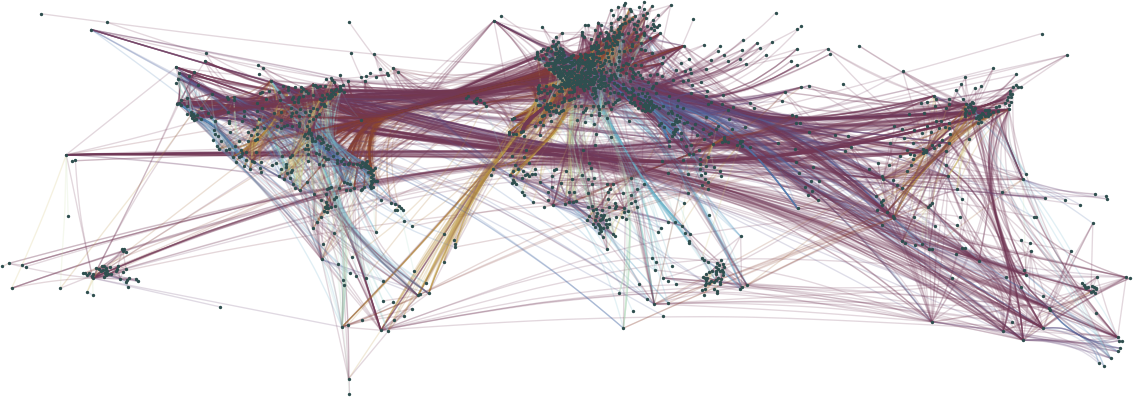}}
        \subfigure[CUBu]{\label{fig:airtrafficKDEEB}\includegraphics[width=0.47\linewidth]{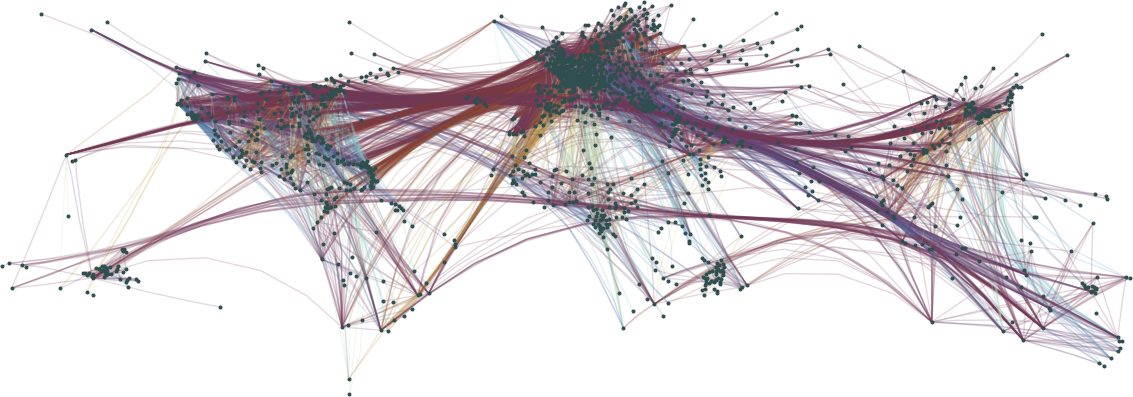}}
        \subfigure[Winding Roads]{\label{fig:airtrafficWR}\includegraphics[width=0.47\linewidth]{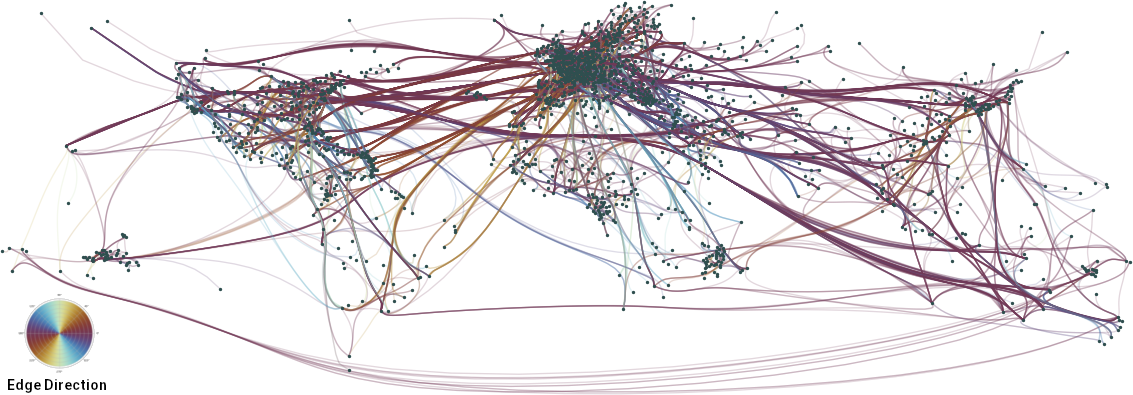}}
        \subfigure[Edge-Path]{\label{fig:airtrafficEBP}\includegraphics[width=0.47\linewidth]{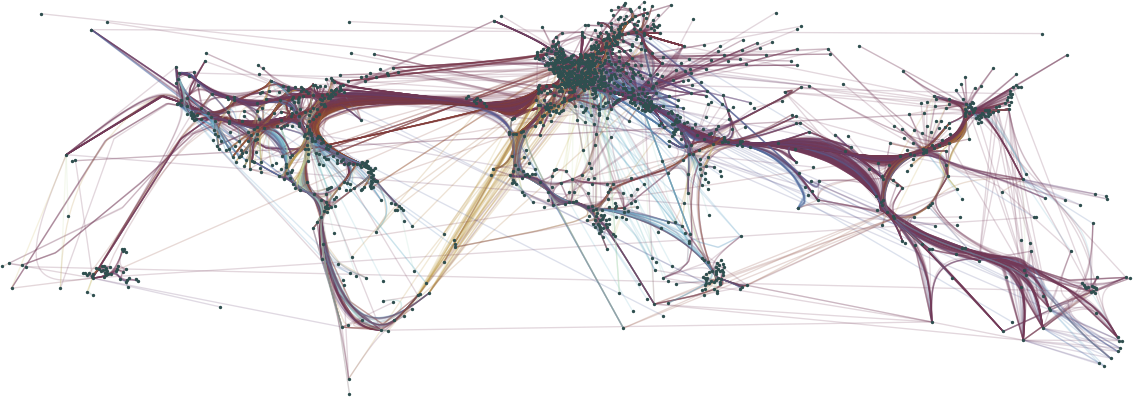}}

    \caption{Results of four algorithms on the Air Traffic network.  {\bf (a)} Force-Directed bundling is able to recover the major trajectories but does not strongly bundle the network.  {\bf (b)} CUBu strongly bundles the main flows of airtraffic, but can suffer from overaggregation.  {\c (c)} Winding Roads divides the traffic into many smaller bundles, but this may not be reflective of underlying graph structure.  {\bf (d)} With Edge-Path bundling, each bundle necessarily reflects a path in the network.  There are separate flows across the Atlantic and Asia that correspond to paths through the network.}
    \label{fig:airtrafficUD}
\end{figure*}

Tables~\ref{tab:real} and~\ref{tab:realDir} show the quality metrics for the real world data and Fig.~\ref{fig:airlinesUndirected} through Fig.~\ref{fig:amazon} show the bundled layouts for this data.

\begin{table*}[t]
\small
    \caption{Scores of the quality metrics for the undirected real-world datasets and all bundling algorithms. Column $\mathrm{dist}$ gives mean and median. Columns $\mathrm{amb}^\delta$ are only shown for $1 \le \delta \le 5$ if there are non-zero entries. Bold values highlight the best score in each column.}
    \centering
    \setlength{\tabcolsep}{1.2ex}
    \begin{tabular}{r||c|cc|c|c||c|cc|c|c|c|c|c||c|cc|c|c|c}
                            & \multicolumn{5}{c||}{US Airlines}   & \multicolumn{8}{c||}{Migrations} & \multicolumn{6}{c}{Air Traffic}  \\
                            &  $\mathrm{ink}_J$   &  \multicolumn{2}{c|}{$\mathrm{dist}$}   &  $\mathrm{amb}^1$ &  $\mathrm{amb}^2$     &   $\mathrm{ink}_J$   &  \multicolumn{2}{c|}{$\mathrm{dist}$}   &  $\mathrm{amb}^1$   &  $\mathrm{amb}^2$ &  $\mathrm{amb}^3$ & $\mathrm{amb}^4$ &  $\mathrm{amb}^5$ &   $\mathrm{ink}_J$   &  \multicolumn{2}{c|}{$\mathrm{dist}$}   &  $\mathrm{amb}^1$   &  $\mathrm{amb}^2$ &  $\mathrm{amb}^3$ \\
                            \hline
        Straight-Line          &  1.00    &  1.00 & 1.00 &  0.66 & 0.02   & 1.00 & 1.00 & 1.00 & 0.71 & 0.25 & 0.06 & 0.02 & 0.01         &    1.00 &  1.00& 1.00 &  0.74 & 0.06 & 0.00 \\
        \hline
        Force      &        0.76    & \textbf{1.03} & \textbf{1.01} &  \textbf{0.85} & 0.03                  & 0.77 & 1.10 & 1.06 & 0.88 & 0.34 & 0.10 & 0.04 & 0.02    &   0.79 &  1.05& 1.02 &  \textbf{0.78} & \textbf{0.08} & \textbf{0.00}  \\
        Confluent           &     0.81    &  1.29 & 1.12 &  0.88  & \textbf{0.02}       & 0.89 & 1.40 & 1.14 & 0.90 & 0.31 & 0.07 & 0.03 & 0.02     &   0.85 &  1.47& 1.28 &  0.90 & 0.09 & \textbf{0.00} \\
        WindingR       &      0.46    & 1.06 & 1.04 &  0.87 & 0.03              & 0.58 & 1.06 & 1.04 & \textbf{0.87} & 0.34 & 0.10 & 0.04 & 0.03    &       0.57 &  {1.05}& 1.04 &  0.81 & 0.12 & 0.01  \\
        KDEEB               &    \textbf{0.30}&  1.21 & 1.15 &  0.90 & 0.09   & \textbf{0.52} & 1.14 & 1.07 & 0.92 & 0.46 & 0.16 & 0.07 & 0.05        &  \textbf{0.39} &  {1.05}& {1.02} &  0.91 & 0.23 & 0.02 \\
        CUBu                &       0.61     & 1.10 & 1.10 & 0.89 & 0.04          & 0.68 & \textbf{1.05} & \textbf{1.02} & 0.90 & 0.32 & 0.08 & 0.03 & \textbf{0.01}    &  0.70 & \textbf{1.00} & \textbf{1.00} & 0.81 & 0.13 & 0.01\\
        Edge-Path           &  0.56    &  1.08 & 1.05 & 0.87 & 0.04         & 0.54 & 1.07 & 1.03 & 0.89 & \textbf{0.24} & \textbf{0.03} & \textbf{0.01} & \textbf{0.01}     &  0.58 &  1.11& 1.08 &  0.89 & 0.15 & 0.01
    \end{tabular}
    \label{tab:real}
\end{table*}

\begin{table*}[t]
\small
    \caption{Scores of the quality metrics for the directed real-world datasets and the directed bundling algorithms. Column $\mathrm{dist}$ gives mean and median. Columns $\mathrm{amb}^\delta$ are only shown for $1 \le \delta \le 5$ if there are non-zero entries. Bold values highlight the best score in each column.}
    \centering
     \setlength{\tabcolsep}{1.2ex}
    \begin{tabular}{r||c|cc|c|c||c|cc|c|c|c|c|c||c|cc|c|c}
                            & \multicolumn{5}{c||}{US Airlines}   & \multicolumn{8}{c||}{Migrations} & \multicolumn{5}{c}{Air Traffic}  \\
                            & $\mathrm{ink}_J$   &  \multicolumn{2}{c|}{$\mathrm{dist}$}   &  $\mathrm{amb}^1$ &  $\mathrm{amb}^2$     &  $\mathrm{ink}_J$   &  \multicolumn{2}{c|}{$\mathrm{dist}$}   &  $\mathrm{amb}^1$   &  $\mathrm{amb}^2$ &  $\mathrm{amb}^3$ & $\mathrm{amb}^4$ &  $\mathrm{amb}^5$ &  $\mathrm{ink}_J$   &  \multicolumn{2}{c|}{$\mathrm{dist}$}   &  $\mathrm{amb}^1$   &  $\mathrm{amb}^2$  \\
                            \hline
        Straight-Line  &    1.00    &  1.00& 1.00 &  0.71 & 0.02   &  1.00 &  1.00 & 1.00 &  0.71 & 0.25 & 0.06 & 0.02 & 0.01           &   1.00 &  1.00 & 1.00 &  0.74 & 0.06 \\
        \hline
        Divided         &    0.75 &  1.08& 1.04 &  0.87 & 0.03               & 0.79 &  1.11 & 1.06 &  \textbf{0.89} & 0.34 & 0.10 & 0.03 & 0.02  & 0.84 &  1.41 & 1.39 &  0.86 & 0.11  \\
        CUBu            &    \textbf{0.58} & \textbf{1.06} & 1.04 & 0.88 & 0.03                 &      0.62 & \textbf{1.06} & \textbf{1.02} & 0.91 & 0.33 & 0.08 & 0.03 & 0.02             & \textbf{0.75} & \textbf{1.00} & \textbf{1.00} & \textbf{0.79} & \textbf{0.09} \\
        Edge-Path       &   0.81 &  1.07& \textbf{1.02} & \textbf{0.83} & \textbf{0.01}       &  \textbf{0.58} &  1.08& 1.04 &  0.90 & \textbf{0.25} & \textbf{0.03} & \textbf{0.01} & \textbf{0.01}   & 0.76 &  1.09& 1.05 &  0.86 & 0.12 
    \end{tabular}
    \label{tab:realDir}
\end{table*}

\subsubsection{Airlines}

\begin{figure*}[p]
    \centering
      \subfigure[Straight Line]{\label{fig:airSL}\includegraphics[width=0.32\linewidth]{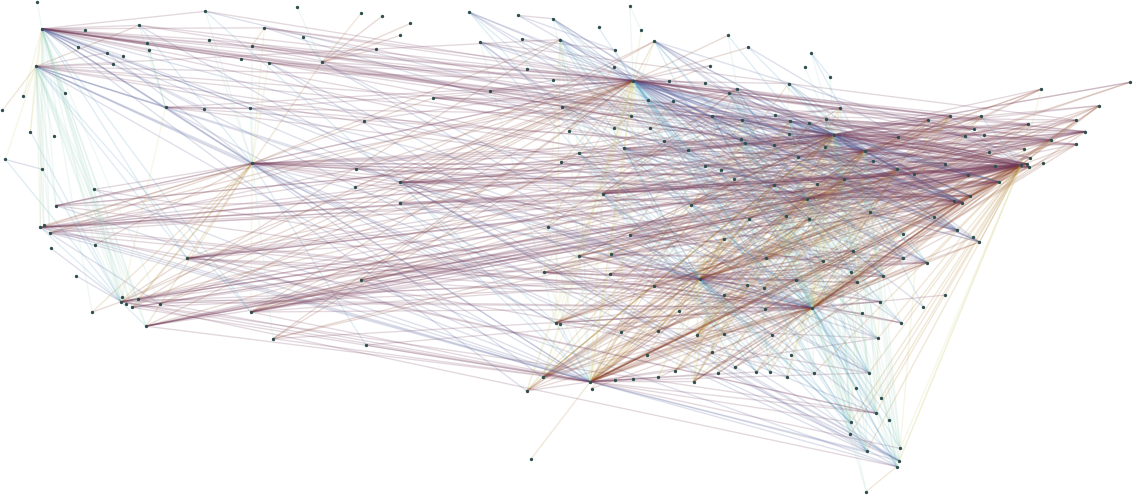}}
      \subfigure[Force-Directed]{\label{fig:airFB}\includegraphics[width=0.32\linewidth]{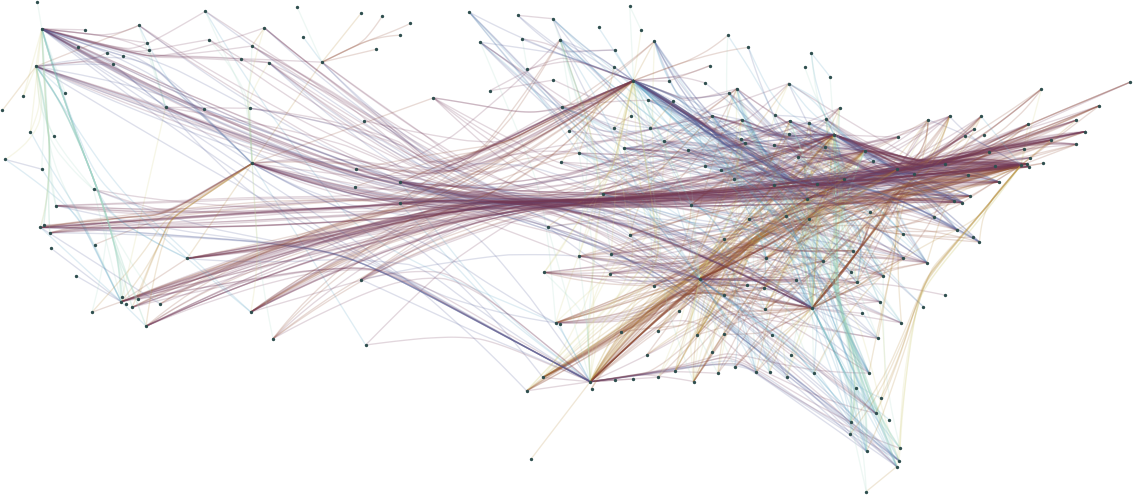}}
      \subfigure[CUBu]{\label{fig:airKDE}\includegraphics[width=0.32\linewidth]{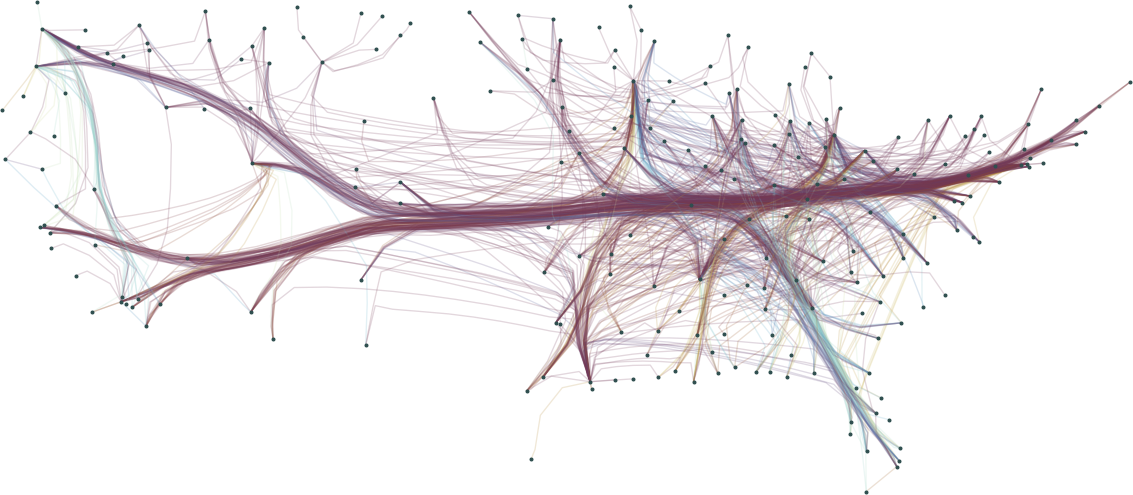}}
      \subfigure[Confluent]{\label{fig:airConf}\includegraphics[width=0.32\linewidth]{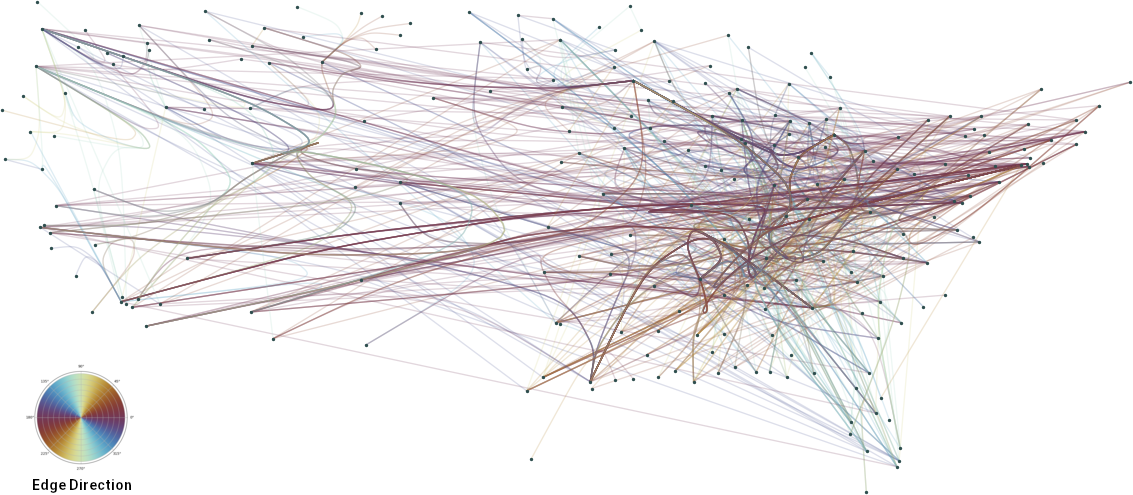}}
      \subfigure[Winding Roads]{\label{fig:airWR}\includegraphics[width=0.32\linewidth]{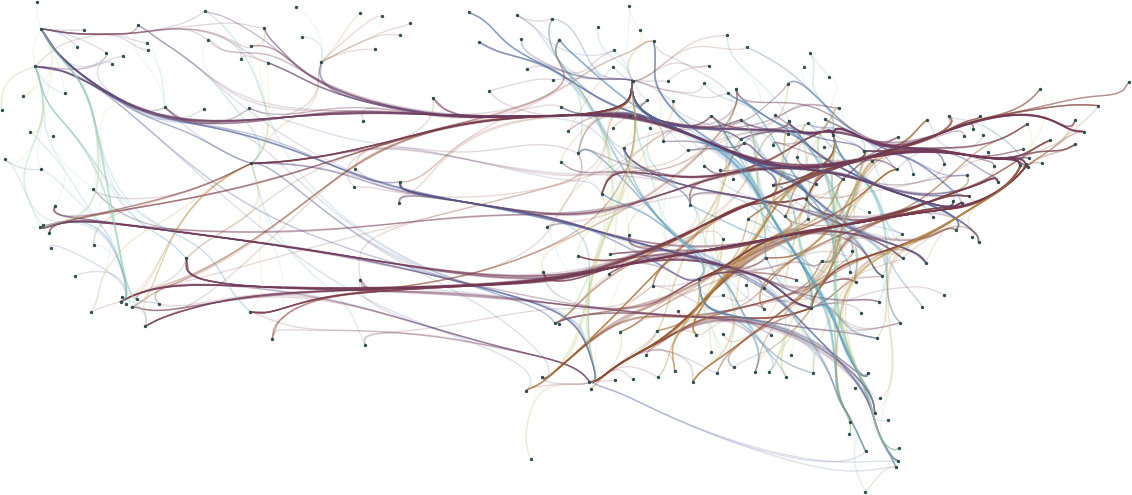}}
      \subfigure[Edge-Path]{\label{fig:airFPB}\includegraphics[width=0.32\linewidth]{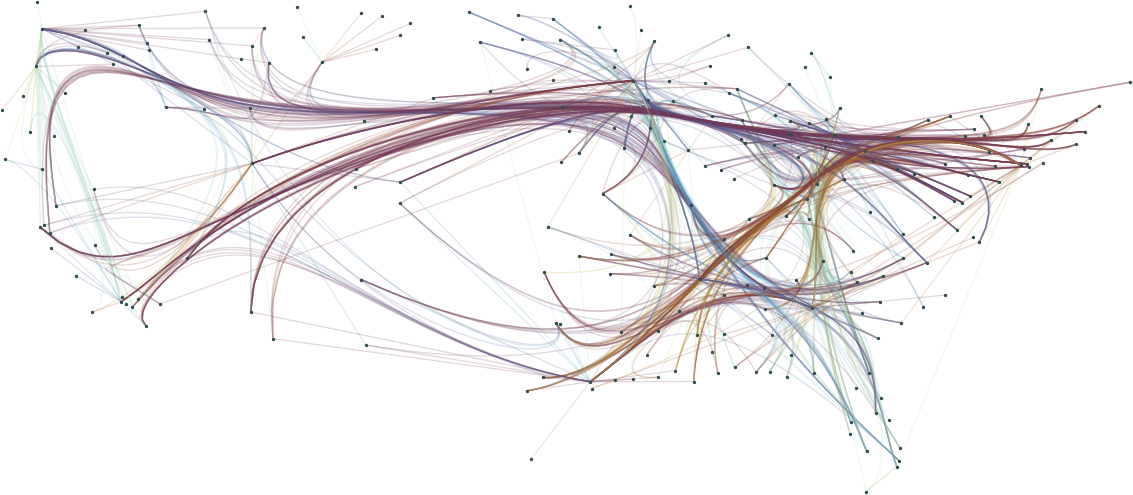}}
    \caption{Airlines (undirected). {\bf (a)} Input drawing.  {\bf(b)} Force-Directed bundling is able to cluster edges into the major flows, but some overaggregation prevents details from being visualised.  {\bf (c)} CUBu provides a good bundling, but also has overaggregation.  {\bf(c)} Confluent drawings can be imposed on the layout, but as the approach cannot layout the graph bicliques can be distantly located, resulting in suboptimal performance.  {\bf(e)} Winding Roads divides the flows into many streams, but these streams can be unfaithful to graph structure.  {\bf (f)} Edge-Path bundling aggregates edges using weighted paths.  The four prominent bundle intersections on the east coast correspond to major airports: Atlanta, Detroit, Minneapolis, and Dallas.}
    \label{fig:airlinesUndirected}
\end{figure*}

{\bf Ink Reduction.}  On the (undirected) Airlines dataset, KDEEB achieves the largest ink reduction followed by Winding Roads, CUBu, and Edge-Path. Force-Directed and Confluent achieve the least ink reduction. These scores also match the visual impression of Fig.~\ref{fig:airlinesUndirected}. 
The ink reduction result would be expected as Edge-Path bundling provides a good compromise between ambiguity and visual simplification.

{\bf Distortion.} Force-Directed performs best with Winding Roads and Edge-Path follow in second. For KDEEB, CUBu, and Confluent we observe higher distortions, which is expected due to the strong ink reduction; for Confluent, as stated before, the bundle routing had not been originally designed for pre-embedded graphs.

{\bf Ambiguity.} The straight-line layout has an ambiguity score of 0.59 due to shallow edge crossings. Force-Directed achieves the least additional ambiguity, followed by Winding Roads and Edge-Path. KDEEB and CUBu are slightly more ambiguous and the poor edge routing in Confluent lets it score highest on ambiguity. This indicates that Confluent, although having good theoretical ambiguity properties, is not well suited if the vertex positions cannot be adjusted.  As $\delta$ is increased, all approaches perform similarly and go to zero when $\delta = 3$.

\hedit{Winding Roads and Edge-Path provide similar scores and minimise overaggregation. Winding Roads produces smaller bundles, prone to fewer shallow crossings, but these bundles are not necessarily backed by structural paths.  Edge-Path bundling produces fewer larger bundles, all of which are backed by graph structure.}

{\bf Qualitative.} Fig.~\ref{fig:airlinesUndirected} shows the results for undirected bundling.  Images for directed bundling are available in the supplementary material. The undirected drawings of Force-Directed, CUBu and KDEEB can clarify coarse structure in the drawing.  In particular, there is a large bundle headed east-west and some of the structure in the densely populated east coast.  Winding Roads is able to divide the large bundles into smaller ones but these do not necessarily correspond to graph structure.  Edge-Path bundling clarifies the two distinct streams on the west coast follow different paths in the network eastward with Denver seeming to act as a control point.  Also, the bundles on the east coast are four distinct path bundles.  These path bundles loosely correspond to four airport hubs:  Atlanta, Minneapolis, Detroit, and Dallas.  %

\subsubsection{Migrations}

{\bf Ink Reduction.}  For the Migrations dataset (Fig.~\ref{fig:teaser}), KDEEB achieves the best ink reduction, closely followed by Edge-Path and then Winding Roads. CUBu and then Force-Directed are next and finally Confluent. These results match the visual impression of the bundled layouts and are also reflected in the directed Migrations data.

{\bf Distortion.} CUBu, Winding Roads, and Edge-Path have the lowest mean distortion scores but very similar medians. Force-Directed and KDEEB follow next. Confluent, as before, scores worst in the distortion metric, which we again attribute to its need to compute a layout simultaneously. These results match our expectations and are similarly observed in the directed Migrations data.

{\bf Ambiguity.}  \hedit{The Migrations Straight-Line layout has a high baseline ambiguity of 0.7. All methods produce similar ambiguity scores for $\delta=1$.  Thus, at this value of $\delta$, shallow edge crossings dominate and all are equally ambiguous. However, as $\delta$ grows, shallow edge crossings matter less in the measure and bundling disconnected edges increase in importance. We show $\delta=1\dots5$ and almost immediately, Edge-Path bundling outperforms all approaches.  It is competitive with Straight-Line as it is able to pull edges connected at small distances away from edges that are not necessarily connected at small values of $\delta$.}

\hedit{On the Migrations dataset Edge-Path bundling is a clear winner with high ink reduction, low distortion, and low ambiguity competitive with straight line drawings.}

{\bf Qualitative.} Fig.~\ref{fig:teaser} shows the undirected bundling approaches on Migrations.  Force-Directed and CUBu 
\hedit{can present}
the main east-west direction of flow and the graph structure on the east coast to some degree, but 
\hedit{aggregate}
to a very high degree and can include unrelated edges.  Winding Roads further subdivides these bundles, but again some of the structure can be obfuscated by clustering unrelated edges together.  Edge-Path only shows flows that are in the underlying graph which are not seen in the other drawings.  The east-west flows actually correspond to two distinct paths:  one 
\hedit{heading}
northward and one 
\hedit{heading}
towards Texas.  On the east coast, the flows are further divided into a smaller number of compact streams that reflect paths in the data.

Table~\ref{tab:realDir} shows the metrics for directed Migrations with similar results.  Fig.~\ref{fig:migrationsDirected} compares directed edge bundling using a force-Directed approach to our own.  The force directed approach is able to divide out streams heading in different directions as can be seen in the east-west parallel flows across the country and the north-south flows on the east and west coast.  However, these patterns contain a number of unrelated edges.  In Edge-Path bundling, the east-west flow divides into three parts around Texas with the thickest bundle heading towards the great lakes region.  On the east coast, more detail is revealed in the north-south direction.  All of this detail depends on the structure of the underlying graph as unrelated edges are not bundled.

\subsubsection{Air Traffic}

{\bf Ink Reduction.}  KDEEB performs the best with Winding Roads and Edge-Path in second.  There is a large gap between the other approaches and these three algorithms.   On the directed datasets, Edge-Path and CUBu outperform force-Directed.  Edge-Path bundling aims at bundling well while reducing the amount of ambiguity in the network and therefore it is competitive with top bundling approaches. 

{\bf Distortion.}  Edge-Path bundling ranks fourth only outperforming confluent, but when looking at the median distortion it performs the same as all other approaches.  Confluent does not perform well on distortion as it is forced to retain the input layout meaning that bicliques could be separated by large distances.  Therefore, on distortion, Edge-Path performs similarly to all approaches.  For directed datasets, Edge-Path and CUBu outperform Divided.

{\bf Ambiguity.}  For $\delta = 1$, Force-Directed has the lowest ambiguity with Winding Roads and CUBu in second, followed by Edge-Path.  There is less bundling in the force-Directed approach, meaning that it has lower ambiguities.  The bundles in Winding Roads are smaller, giving it an advantage in terms of this metric.  KDEEB and confluent are not too far away but slightly more ambiguous.  With increasing $\delta$, the approaches become comparable rapidly.  In the directed case, CUBu and Edge-Path perform similarly with similar findings.

{\bf Qualitative.}  Fig.~\ref{fig:airtrafficUD} shows the results of Force-Directed, CUBu, Winding Roads, and Edge-Path bundling on this global airline dataset.  Force-Directed is able to simplify the east-west flows across the Atlantic and Pacific, but only at a high level with many of the edges remaining unaggregated.  CUBu is able to simplify these flows, but suffers from some overaggregation.  Winding Roads divides these flows into several smaller streams, but these streams may not be reflective of graph structure.  Edge-Path bundling provides a compromise between level of bundling and ambiguity.  In this image, flows from Europe to South America are a separate yellow bundle, indicating few independent paths.  Flights directly across the Atlantic are divided into separate bundles that represent paths in the graph as well as detail in the United States being revealed.  Flights in Asia split into a triangle of streams towards Australia with north-south flights being in dependant from flights on the continent.  Edge-Path bundling is able to reveal structure in this network that is more faithful to the graph structure.   

\begin{figure}
    \centering
        \subfigure[Edge-Path]{\label{fig:amazonEbp}\includegraphics[width=0.44\linewidth]{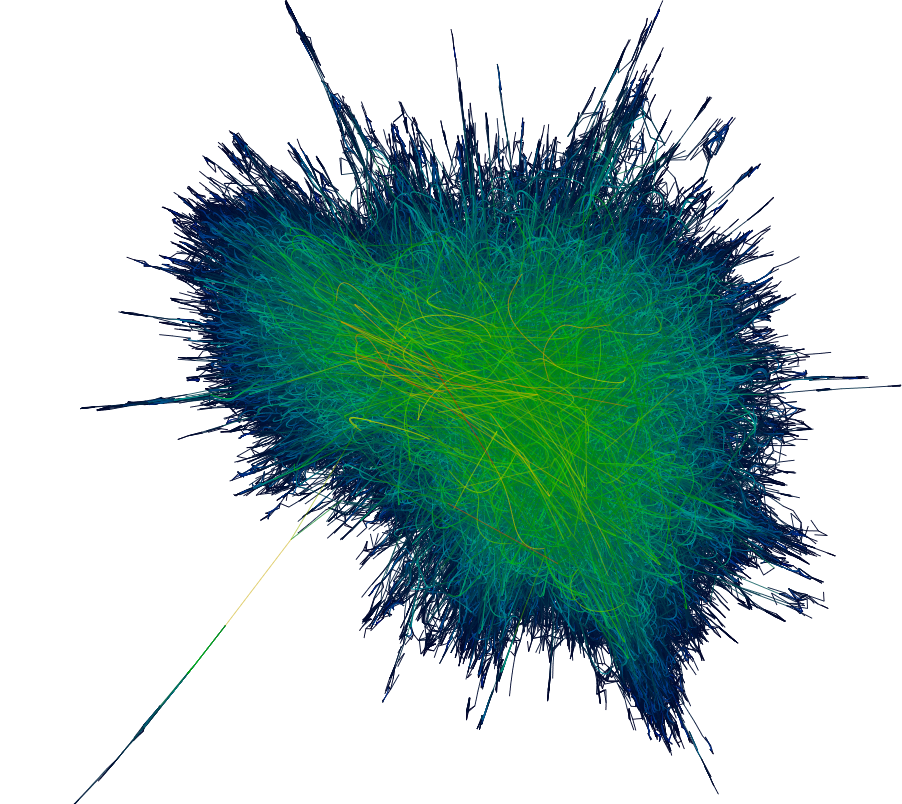}}
        \subfigure[CUBu]{\label{fig:amazonCubu}\includegraphics[width=0.44\linewidth]{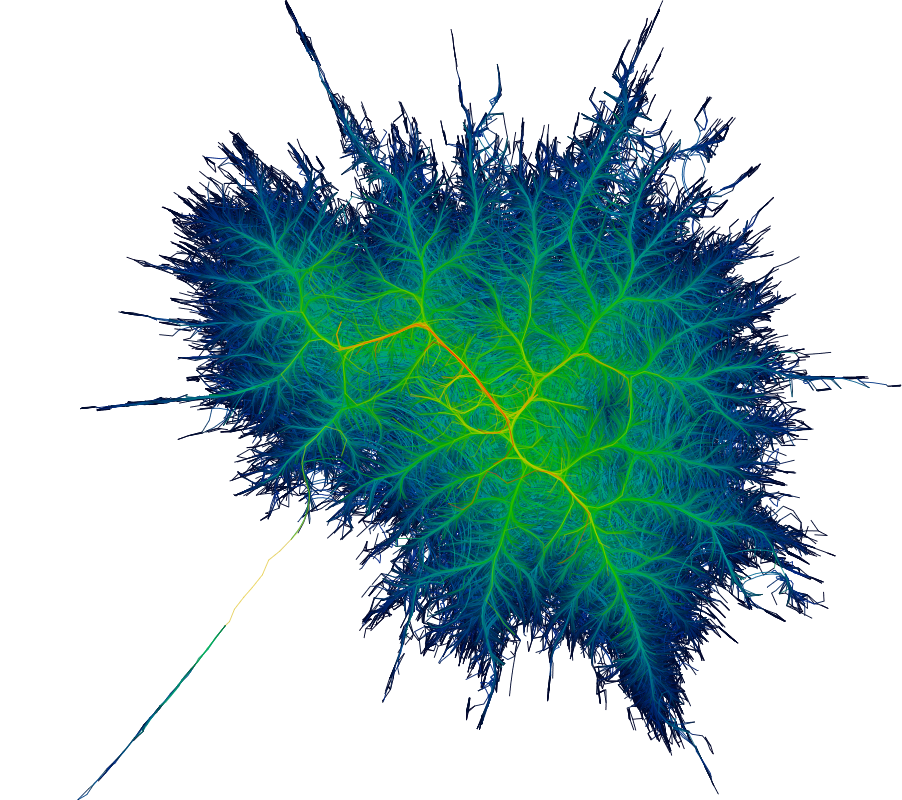}}
    \caption{Edge-Path bundling and CUBu on Amazon Subset.  CUBu rendering style is used on both drawings.}
    \label{fig:amazon}
\end{figure}

\subsubsection{Amazon Subset}

Fig.~\ref{fig:amazon} shows the Amazon Subset using the CUBu rendering style.  Images of other algorithms that were able to bundle the graph are in the supplementary material. The dataset was too large for our ambiguity metric, so metric values are not reported. CUBu bundling renders a tree-like pattern whereas Edge-Path does not. Edge-Path bundling takes into account the graph structure (faithfulness~\cite{NguyenEH17}). CUBu has no such guarantee, because it does not take this structure into account. However, the tree-like structure emerges from bundling edges with similar attributes together such as proximity and direction.

\section{Discussion, Limitations, and Conclusion}

Edge-Path bundling is not without its ambiguities even though it completely eliminates independent edge ambiguities.  As shown in Fig.~\ref{fig:pathamb}, direct connections between vertices can be ambiguous, but paths between all vertices in the cluster will exist by definition.  This edge aggregation is similar to node aggregation strategies~\cite{archambault_2008,archambault_2011,golodetz_2020,vanHam_2004,abello_2006,batagelj_2011}.  In visualisations produced by node aggregation strategies, it is impossible to understand \textit{how} the vertices within the aggregated node connect, only that they are connected; in Edge-Path bundling, we know the bundle is connected and paths exist between all vertices, but specific connections are difficult to read.  Interactive techniques, similar to Edgelens~\cite{wong_2003}, could be explored to disambiguate this ambiguity. Edge-Path bundles can be more convoluted as they must follow shortest paths in a graph.  %
Detection of these loops would solve this issue but remains future work. 

When bundling directed graphs, Edge-Path bundling can give poor results, 
since
a directed graph may have many forward and backward edges between 
vertex pairs.
This can lead to edges being bundled along the same path in opposite directions. When combined with overplotting, this may result in indistinguishable bundles. Future work could focus on a better routing approach for bundles or avoiding to use the same path in both directions \hedit{or using a parallel bundling style~\cite{vdzwan_2016}}.  Edge-Path bundling takes a graph and a layout as input to compute the bundled drawing. 
Interesting
future work would be to design layout algorithms that maximise the Edge-Path bundling as part of the layout process.

\hedit{Our implementation of Edge-Path bundling has a worst-case complexity of $O(|E|^2\log |V|)$, but still works well on graphs of up to ten thousand edges.  For dense networks where $|E|$ approaches $|V|^2$, the performance deteriorates. From a theoretical standpoint, finding approximations or algorithms that are faster are of interest. Multilevel bundling~\cite{gansner_2011}, faster shortest path implementations~\cite{bdgmps-rptn-16} or parallelisation~\cite{ZhangLGJHLH20} could be applied to improve the practical performance}.
\hedit{Similarly, the scalability of the ambiguity metric is an open problem}.

We presented a method for bundling graphs by considering Edge-Path pairs. The resulting drawings do not incur independent edge ambiguities and are more resilient to creating signal through bundling when there is none.  Our resulting bundles provide a new compromise between degree of bundling and faithfulness to the  graph structure.

\acknowledgments{
We acknowledge funding by the Vienna Science and Technology Fund (WWTF) through project ICT19-035, UKRI EP/V033670/1
and Academy of Finland grant 327352.}

\bibliographystyle{abbrv-doi-hyperref}

\bibliography{bibliography}

\begin{thebibliography}{10}

\bibitem{Telea2021Cubu}
\href{https://webspace.science.uu.nl/~telea001/InfoVis/CUBu}{{CUBu -
  Implementation}}.
\newblock
  \href{https://webspace.science.uu.nl/~telea001/InfoVis/CUBu}{\url{https://webspace.science.uu.nl/~telea001/InfoVis/CUBu}}.
\newblock \href{https://webspace.science.uu.nl/~telea001/InfoVis/CUBu}{[Online;
  accessed 5. Aug. 2021]}.

\bibitem{upphiminn2021Aug}
\href{https://github.com/upphiminn/d3.ForceBundle}{{d3.ForceBundle - Github}}.
\newblock
  \href{https://github.com/upphiminn/d3.ForceBundle}{\url{https://github.com/upphiminn/d3.ForceBundle}}.
\newblock \href{https://github.com/upphiminn/d3.ForceBundle}{[Online; accessed
  5. Aug. 2021]}.

\bibitem{kakearney2021Aug}
\href{https://github.com/kakearney/divedgebundle-pkg}{{Divided Edge Bundling -
  Implementation}}.
\newblock
  \href{https://github.com/kakearney/divedgebundle-pkg}{\url{https://github.com/kakearney/divedgebundle-pk}}.
\newblock \href{https://github.com/kakearney/divedgebundle-pkg}{[Online;
  accessed 5. Aug. 2021]}.

\bibitem{Hurter2019Jul}
\href{http://recherche.enac.fr/~hurter/KDEEB.html}{{KDEEB - Implementation}}.
\newblock
  \href{http://recherche.enac.fr/~hurter/KDEEB.html}{\url{http://recherche.enac.fr/~hurter/KDEEB.html}}.
\newblock \href{http://recherche.enac.fr/~hurter/KDEEB.html}{[Online; accessed
  5. Aug. 2021]}.

\bibitem{jxz122021Aug}
\href{https://github.com/jxz12/pconfluent}{{pconfluent - Github}}.
\newblock
  \href{https://github.com/jxz12/pconfluent}{\url{https://github.com/jxz12/pconfluent}}.
\newblock \href{https://github.com/jxz12/pconfluent}{[Online; accessed 5. Aug.
  2021]}.

\bibitem{abello_2006}
\href{https://doi.org/10.1109/TVCG.2006.120}{J.~{Abello}, F.~{van Ham}, and
  N.~{Krishnan}}.
\newblock \href{https://doi.org/10.1109/TVCG.2006.120}{{ASK-GraphView:} a large
  scale graph visualization system}.
\newblock \href{https://doi.org/10.1109/TVCG.2006.120}{{\em IEEE Trans.
  Visualization and Computer Graphics}},
  \href{https://doi.org/10.1109/TVCG.2006.120}{12(5):669--676},
  \href{https://doi.org/10.1109/TVCG.2006.120}{2006}.
  \href{https://doi.org/10.1109/TVCG.2006.120}
{doi: {{%
10\hspace{.1pt}\discretionary{.}{%
}{.}\hspace{.4pt}1109\discretionary{/}{%
}{/}TVCG\hspace{.1pt}\discretionary{.}{%
}{.}\hspace{.4pt}2006\hspace{.1pt}\discretionary{.}{%
}{.}\hspace{.4pt}120}}}


\bibitem{archambault_2008}
\href{https://doi.org/10.1109/TVCG.2008.34}{D.~{Archambault}, T.~{Munzner}, and
  D.~{Auber}}.
\newblock \href{https://doi.org/10.1109/TVCG.2008.34}{{GrouseFlocks}: Steerable
  exploration of graph hierarchy space}.
\newblock \href{https://doi.org/10.1109/TVCG.2008.34}{{\em IEEE Trans.
  Visualization and Computer Graphics}},
  \href{https://doi.org/10.1109/TVCG.2008.34}{14(4):900--913},
  \href{https://doi.org/10.1109/TVCG.2008.34}{2008}.
  \href{https://doi.org/10.1109/TVCG.2008.34}
{doi: {{%
10\hspace{.1pt}\discretionary{.}{%
}{.}\hspace{.4pt}1109\discretionary{/}{%
}{/}TVCG\hspace{.1pt}\discretionary{.}{%
}{.}\hspace{.4pt}2008\hspace{.1pt}\discretionary{.}{%
}{.}\hspace{.4pt}34}}}


\bibitem{archambault_2011}
\href{https://doi.org/10.1109/TVCG.2010.60}{D.~{Archambault}, T.~{Munzner}, and
  D.~{Auber}}.
\newblock \href{https://doi.org/10.1109/TVCG.2010.60}{Tugging graphs faster:
  Efficiently modifying path-preserving hierarchies for browsing paths}.
\newblock \href{https://doi.org/10.1109/TVCG.2010.60}{{\em IEEE Trans.
  Visualization and Computer Graphics}},
  \href{https://doi.org/10.1109/TVCG.2010.60}{17(3):276--289},
  \href{https://doi.org/10.1109/TVCG.2010.60}{2011}.
  \href{https://doi.org/10.1109/TVCG.2010.60}
{doi: {{%
10\hspace{.1pt}\discretionary{.}{%
}{.}\hspace{.4pt}1109\discretionary{/}{%
}{/}TVCG\hspace{.1pt}\discretionary{.}{%
}{.}\hspace{.4pt}2010\hspace{.1pt}\discretionary{.}{%
}{.}\hspace{.4pt}60}}}


\bibitem{auber_2018}
\href{https://doi.org/10.1007/978-1-4939-7131-2_315}{D.~Auber, D.~Archambault,
  R.~Bourqui, M.~Delest, J.~Dubois, A.~Lambert, P.~Mary, M.~Mathiaut,
  G.~Melan{\c{c}}on, B.~Pinaud, B.~Renoust, and J.~Vallet}.
\newblock \href{https://doi.org/10.1007/978-1-4939-7131-2_315}{Tulip 5}.
\newblock \href{https://doi.org/10.1007/978-1-4939-7131-2_315}{In R.~Alhajj and
  J.~Rokne, eds., {\em Encyclopedia of Social Network Analysis and Mining}},
  \href{https://doi.org/10.1007/978-1-4939-7131-2_315}{pp. 3185--3212}.
  \href{https://doi.org/10.1007/978-1-4939-7131-2_315}{Springer},
  \href{https://doi.org/10.1007/978-1-4939-7131-2_315}{2018}.
  \href{https://doi.org/10.1007/978-1-4939-7131-2_315}
{doi: {{%
10\hspace{.1pt}\discretionary{.}{%
}{.}\hspace{.4pt}1007\discretionary{/}{%
}{/}978\discretionary{%
}{-}{-}1\discretionary{%
}{-}{-}4939\discretionary{%
}{-}{-}7131\discretionary{%
}{-}{-}2\_315}}}


\bibitem{bach_2016}
\href{https://doi.org/10.1109/TVCG.2016.2598958}{B.~{Bach}, N.~H. {Riche},
  C.~{Hurter}, K.~{Marriott}, and T.~{Dwyer}}.
\newblock \href{https://doi.org/10.1109/TVCG.2016.2598958}{Towards unambiguous
  edge bundling: Investigating confluent drawings for network visualization}.
\newblock \href{https://doi.org/10.1109/TVCG.2016.2598958}{{\em IEEE Trans.
  Visualization and Computer Graphics}},
  \href{https://doi.org/10.1109/TVCG.2016.2598958}{23(1):541--550},
  \href{https://doi.org/10.1109/TVCG.2016.2598958}{2017}.
  \href{https://doi.org/10.1109/TVCG.2016.2598958}
{doi: {{%
10\hspace{.1pt}\discretionary{.}{%
}{.}\hspace{.4pt}1109\discretionary{/}{%
}{/}TVCG\hspace{.1pt}\discretionary{.}{%
}{.}\hspace{.4pt}2016\hspace{.1pt}\discretionary{.}{%
}{.}\hspace{.4pt}2598958}}}


\bibitem{bdgmps-rptn-16}
\href{https://doi.org/10.1007/978-3-319-49487-6_2}{H.~Bast, D.~Delling,
  A.~Goldberg, M.~Müller-Hannemann, T.~Pajor, P.~Sanders, D.~Wagner, and R.~F.
  Werneck}.
\newblock \href{https://doi.org/10.1007/978-3-319-49487-6_2}{Route planning in
  transportation networks}.
\newblock \href{https://doi.org/10.1007/978-3-319-49487-6_2}{In L.~Kliemann and
  P.~Sanders, eds., {\em Algorithm Engineering}},
  \href{https://doi.org/10.1007/978-3-319-49487-6_2}{vol. 9220 of {\em LNCS}},
  \href{https://doi.org/10.1007/978-3-319-49487-6_2}{chap.~2, pp. 19--80}.
  \href{https://doi.org/10.1007/978-3-319-49487-6_2}{Springer},
  \href{https://doi.org/10.1007/978-3-319-49487-6_2}{2016}.
  \href{https://doi.org/10.1007/978-3-319-49487-6_2}
{doi: {{%
10\hspace{.1pt}\discretionary{.}{%
}{.}\hspace{.4pt}1007\discretionary{/}{%
}{/}978\discretionary{%
}{-}{-}3\discretionary{%
}{-}{-}319\discretionary{%
}{-}{-}49487\discretionary{%
}{-}{-}6\_2}}}


\bibitem{batagelj_2011}
\href{https://doi.org/10.1109/TVCG.2010.265}{V.~{Batagelj}, F.~J.
  {Brandenburg}, W.~{Didimo}, G.~{Liotta}, P.~{Palladino}, and
  M.~{Patrignani}}.
\newblock \href{https://doi.org/10.1109/TVCG.2010.265}{Visual analysis of large
  graphs using (x,y)-clustering and hybrid visualizations}.
\newblock \href{https://doi.org/10.1109/TVCG.2010.265}{{\em IEEE Trans.
  Visualization and Computer Graphics}},
  \href{https://doi.org/10.1109/TVCG.2010.265}{17(11):1587--1598},
  \href{https://doi.org/10.1109/TVCG.2010.265}{2011}.
  \href{https://doi.org/10.1109/TVCG.2010.265}
{doi: {{%
10\hspace{.1pt}\discretionary{.}{%
}{.}\hspace{.4pt}1109\discretionary{/}{%
}{/}TVCG\hspace{.1pt}\discretionary{.}{%
}{.}\hspace{.4pt}2010\hspace{.1pt}\discretionary{.}{%
}{.}\hspace{.4pt}265}}}


\bibitem{bruls_2000}
\href{https://doi.org/10.1007/978-3-7091-6783-0_4}{M.~Bruls, K.~Huizing, and
  J.~J. van Wijk}.
\newblock \href{https://doi.org/10.1007/978-3-7091-6783-0_4}{Squarified
  treemaps}.
\newblock \href{https://doi.org/10.1007/978-3-7091-6783-0_4}{In W.~C. de~Leeuw
  and R.~van Liere, eds., {\em 2nd Joint Eurographics - {IEEE} {TCVG} Symposium
  on Visualization}}, \href{https://doi.org/10.1007/978-3-7091-6783-0_4}{pp.
  33--42}. \href{https://doi.org/10.1007/978-3-7091-6783-0_4}{Eurographics
  Association}, \href{https://doi.org/10.1007/978-3-7091-6783-0_4}{2000}.
  \href{https://doi.org/10.1007/978-3-7091-6783-0_4}
{doi: {{%
10\hspace{.1pt}\discretionary{.}{%
}{.}\hspace{.4pt}1007\discretionary{/}{%
}{/}978\discretionary{%
}{-}{-}3\discretionary{%
}{-}{-}7091\discretionary{%
}{-}{-}6783\discretionary{%
}{-}{-}0\_4}}}


\bibitem{crameri_2020}
F.~Crameri, G.~E. Shephard, and P.~J. Heron.
\newblock The misuse of colour in science communication.
\newblock {\em Nature communications}, 11(1):1--10, 2020.

\bibitem{cui_2008}
\href{https://doi.org/10.1109/TVCG.2008.135}{W.~{Cui}, H.~{Zhou}, H.~{Qu},
  P.~C. {Wong}, and X.~{Li}}.
\newblock \href{https://doi.org/10.1109/TVCG.2008.135}{Geometry-based edge
  clustering for graph visualization}.
\newblock \href{https://doi.org/10.1109/TVCG.2008.135}{{\em IEEE Trans.
  Visualization and Computer Graphics}},
  \href{https://doi.org/10.1109/TVCG.2008.135}{14(6):1277--1284},
  \href{https://doi.org/10.1109/TVCG.2008.135}{2008}.
  \href{https://doi.org/10.1109/TVCG.2008.135}
{doi: {{%
10\hspace{.1pt}\discretionary{.}{%
}{.}\hspace{.4pt}1109\discretionary{/}{%
}{/}TVCG\hspace{.1pt}\discretionary{.}{%
}{.}\hspace{.4pt}2008\hspace{.1pt}\discretionary{.}{%
}{.}\hspace{.4pt}135}}}


\bibitem{dickerson_2005}
\href{https://doi.org/10.7155/jgaa.00099}{M.~{Dickerson}, D.~{Eppstein}, M.~T.
  {Goodrich}, and J.~Y. {Meng}}.
\newblock \href{https://doi.org/10.7155/jgaa.00099}{Confluent drawings:
  Visualizing non-planar diagrams in a planar way}.
\newblock \href{https://doi.org/10.7155/jgaa.00099}{{\em J. Graph Algorithms
  and Applications}}, \href{https://doi.org/10.7155/jgaa.00099}{9(1):31--52},
  \href{https://doi.org/10.7155/jgaa.00099}{2005}.
  \href{https://doi.org/10.7155/jgaa.00099}
{doi: {{%
10\hspace{.1pt}\discretionary{.}{%
}{.}\hspace{.4pt}7155\discretionary{/}{%
}{/}jgaa\hspace{.1pt}\discretionary{.}{%
}{.}\hspace{.4pt}00099}}}


\bibitem{egm-dd-06}
\href{https://doi.org/10.1007/11618058_16}{D.~Eppstein, M.~T. Goodrich, and
  J.~Y. Meng}.
\newblock \href{https://doi.org/10.1007/11618058_16}{Delta-confluent drawings}.
\newblock \href{https://doi.org/10.1007/11618058_16}{In P.~Healy and
  N.~Nikolov, eds., {\em Graph Drawing (GD'05)}},
  \href{https://doi.org/10.1007/11618058_16}{vol. 3843 of {\em LNCS}},
  \href{https://doi.org/10.1007/11618058_16}{pp. 165--176}.
  \href{https://doi.org/10.1007/11618058_16}{Springer},
  \href{https://doi.org/10.1007/11618058_16}{2006}.
  \href{https://doi.org/10.1007/11618058_16}
{doi: {{%
10\hspace{.1pt}\discretionary{.}{%
}{.}\hspace{.4pt}1007\discretionary{/}{%
}{/}11618058\_16}}}


\bibitem{ehlnsv-scd-16}
\href{https://doi.org/10.20382/jocg.v7i1a2}{D.~Eppstein, D.~Holten,
  M.~L{\"o}ffler, M.~N{\"o}llenburg, B.~Speckmann, and K.~Verbeek}.
\newblock \href{https://doi.org/10.20382/jocg.v7i1a2}{Strict confluent
  drawing}.
\newblock \href{https://doi.org/10.20382/jocg.v7i1a2}{{\em J. Computational
  Geometry}}, \href{https://doi.org/10.20382/jocg.v7i1a2}{7(1):22--46},
  \href{https://doi.org/10.20382/jocg.v7i1a2}{2016}.
  \href{https://doi.org/10.20382/jocg.v7i1a2}
{doi: {{%
10\hspace{.1pt}\discretionary{.}{%
}{.}\hspace{.4pt}20382\discretionary{/}{%
}{/}jocg\hspace{.1pt}\discretionary{.}{%
}{.}\hspace{.4pt}v7i1a2}}}


\bibitem{ersoy_2011}
\href{https://doi.org/10.1109/TVCG.2011.233}{O.~{Ersoy}, C.~{Hurter},
  F.~{Paulovich}, G.~{Cantareiro}, and A.~{Telea}}.
\newblock \href{https://doi.org/10.1109/TVCG.2011.233}{Skeleton-based edge
  bundling for graph visualization}.
\newblock \href{https://doi.org/10.1109/TVCG.2011.233}{{\em IEEE Trans.
  Visualization and Computer Graphics}},
  \href{https://doi.org/10.1109/TVCG.2011.233}{17(12):2364--2373},
  \href{https://doi.org/10.1109/TVCG.2011.233}{2011}.
  \href{https://doi.org/10.1109/TVCG.2011.233}
{doi: {{%
10\hspace{.1pt}\discretionary{.}{%
}{.}\hspace{.4pt}1109\discretionary{/}{%
}{/}TVCG\hspace{.1pt}\discretionary{.}{%
}{.}\hspace{.4pt}2011\hspace{.1pt}\discretionary{.}{%
}{.}\hspace{.4pt}233}}}


\bibitem{fgkn-sog-19}
\href{https://doi.org/10.1007/978-3-030-35802-0_12}{H.~F{\"o}rster, R.~Ganian,
  F.~Klute, and M.~N{\"o}llenburg}.
\newblock \href{https://doi.org/10.1007/978-3-030-35802-0_12}{On strict
  (outer-) confluent graphs}.
\newblock \href{https://doi.org/10.1007/978-3-030-35802-0_12}{In D.~Archambault
  and C.~D. T{\'o}th, eds., {\em Graph Drawing and Network Visualization
  (GD'19)}}, \href{https://doi.org/10.1007/978-3-030-35802-0_12}{vol. 11904 of
  {\em LNCS}}, \href{https://doi.org/10.1007/978-3-030-35802-0_12}{pp.
  147--161}. \href{https://doi.org/10.1007/978-3-030-35802-0_12}{Springer},
  \href{https://doi.org/10.1007/978-3-030-35802-0_12}{2019}.
  \href{https://doi.org/10.1007/978-3-030-35802-0_12}
{doi: {{%
10\hspace{.1pt}\discretionary{.}{%
}{.}\hspace{.4pt}1007\discretionary{/}{%
}{/}978\discretionary{%
}{-}{-}3\discretionary{%
}{-}{-}030\discretionary{%
}{-}{-}35802\discretionary{%
}{-}{-}0\_12}}}


\bibitem{gansner_2011}
\href{https://doi.org/10.1109/PACIFICVIS.2011.5742389}{E.~R. {Gansner},
  Y.~{Hu}, S.~{North}, and C.~{Scheidegger}}.
\newblock \href{https://doi.org/10.1109/PACIFICVIS.2011.5742389}{Multilevel
  agglomerative edge bundling for visualizing large graphs}.
\newblock \href{https://doi.org/10.1109/PACIFICVIS.2011.5742389}{In {\em
  Pacific Visualization Symposium (PacificVis '11)}},
  \href{https://doi.org/10.1109/PACIFICVIS.2011.5742389}{pp. 187--194}.
  \href{https://doi.org/10.1109/PACIFICVIS.2011.5742389}{IEEE},
  \href{https://doi.org/10.1109/PACIFICVIS.2011.5742389}{2011}.
  \href{https://doi.org/10.1109/PACIFICVIS.2011.5742389}
{doi: {{%
10\hspace{.1pt}\discretionary{.}{%
}{.}\hspace{.4pt}1109\discretionary{/}{%
}{/}PACIFICVIS\hspace{.1pt}\discretionary{.}{%
}{.}\hspace{.4pt}2011\hspace{.1pt}\discretionary{.}{%
}{.}\hspace{.4pt}5742389}}}


\bibitem{golodetz_2020}
\href{https://doi.org/https://doi.org/10.1016/j.patcog.2020.107257}{S.~Golodetz,
  A.~Arnab, I.~Voiculescu, and S.~Cameron}.
\newblock
  \href{https://doi.org/https://doi.org/10.1016/j.patcog.2020.107257}{Simplifying
  tuggraph using zipping algorithms}.
\newblock
  \href{https://doi.org/https://doi.org/10.1016/j.patcog.2020.107257}{{\em
  Pattern Recognition}},
  \href{https://doi.org/https://doi.org/10.1016/j.patcog.2020.107257}{103:107257},
  \href{https://doi.org/https://doi.org/10.1016/j.patcog.2020.107257}{2020}.
  \href{https://doi.org/10.1016/j.patcog.2020.107257}
{doi: {{%
10\hspace{.1pt}\discretionary{.}{%
}{.}\hspace{.4pt}1016\discretionary{/}{%
}{/}j\hspace{.1pt}\discretionary{.}{%
}{.}\hspace{.4pt}patcog\hspace{.1pt}\discretionary{.}{%
}{.}\hspace{.4pt}2020\hspace{.1pt}\discretionary{.}{%
}{.}\hspace{.4pt}107257}}}


\bibitem{hmr-becgcd-07}
\href{https://doi.org/10.1007/978-3-540-70904-6_39}{M.~Hirsch, H.~Meijer, and
  D.~Rappaport}.
\newblock \href{https://doi.org/10.1007/978-3-540-70904-6_39}{Biclique edge
  cover graphs and confluent drawings}.
\newblock \href{https://doi.org/10.1007/978-3-540-70904-6_39}{In {\em Graph
  Drawing (GD'06)}}, \href{https://doi.org/10.1007/978-3-540-70904-6_39}{vol.
  4372 of {\em LNCS}}, \href{https://doi.org/10.1007/978-3-540-70904-6_39}{pp.
  405--416}. \href{https://doi.org/10.1007/978-3-540-70904-6_39}{Springer},
  \href{https://doi.org/10.1007/978-3-540-70904-6_39}{2007}.
  \href{https://doi.org/10.1007/978-3-540-70904-6_39}
{doi: {{%
10\hspace{.1pt}\discretionary{.}{%
}{.}\hspace{.4pt}1007\discretionary{/}{%
}{/}978\discretionary{%
}{-}{-}3\discretionary{%
}{-}{-}540\discretionary{%
}{-}{-}70904\discretionary{%
}{-}{-}6\_39}}}


\bibitem{holten_2006}
\href{https://doi.org/10.1109/TVCG.2006.147}{D.~{Holten}}.
\newblock \href{https://doi.org/10.1109/TVCG.2006.147}{Hierarchical edge
  bundles: Visualization of adjacency relations in hierarchical data}.
\newblock \href{https://doi.org/10.1109/TVCG.2006.147}{{\em IEEE Trans.
  Visualization and Computer Graphics}},
  \href{https://doi.org/10.1109/TVCG.2006.147}{12(5):741--748},
  \href{https://doi.org/10.1109/TVCG.2006.147}{2006}.
  \href{https://doi.org/10.1109/TVCG.2006.147}
{doi: {{%
10\hspace{.1pt}\discretionary{.}{%
}{.}\hspace{.4pt}1109\discretionary{/}{%
}{/}TVCG\hspace{.1pt}\discretionary{.}{%
}{.}\hspace{.4pt}2006\hspace{.1pt}\discretionary{.}{%
}{.}\hspace{.4pt}147}}}


\bibitem{holten_2009}
\href{https://doi.org/10.1111/j.1467-8659.2009.01450.x}{D.~Holten and J.~J. van
  Wijk}.
\newblock
  \href{https://doi.org/10.1111/j.1467-8659.2009.01450.x}{Force-directed edge
  bundling for graph visualization}.
\newblock \href{https://doi.org/10.1111/j.1467-8659.2009.01450.x}{{\em Computer
  Graphics Forum}},
  \href{https://doi.org/10.1111/j.1467-8659.2009.01450.x}{28(3):983--990},
  \href{https://doi.org/10.1111/j.1467-8659.2009.01450.x}{2009}.
  \href{https://doi.org/10.1111/j.1467-8659.2009.01450.x}
{doi: {{%
10\hspace{.1pt}\discretionary{.}{%
}{.}\hspace{.4pt}1111\discretionary{/}{%
}{/}j\hspace{.1pt}\discretionary{.}{%
}{.}\hspace{.4pt}1467\discretionary{%
}{-}{-}8659\hspace{.1pt}\discretionary{.}{%
}{.}\hspace{.4pt}2009\hspace{.1pt}\discretionary{.}{%
}{.}\hspace{.4pt}01450\hspace{.1pt}\discretionary{.}{%
}{.}\hspace{.4pt}x}}}


\bibitem{heh-grbgt-09}
\href{https://doi.org/10.1109/PACIFICVIS.2009.4906848}{W.~Huang, P.~Eades, and
  S.-H. Hong}.
\newblock \href{https://doi.org/10.1109/PACIFICVIS.2009.4906848}{A graph
  reading behavior: Geodesic-path tendency}.
\newblock \href{https://doi.org/10.1109/PACIFICVIS.2009.4906848}{In {\em
  Pacific Visualization Symposium (PacificVis'09)}},
  \href{https://doi.org/10.1109/PACIFICVIS.2009.4906848}{pp. 137--144}.
  \href{https://doi.org/10.1109/PACIFICVIS.2009.4906848}{IEEE},
  \href{https://doi.org/10.1109/PACIFICVIS.2009.4906848}{2009}.
  \href{https://doi.org/10.1109/PACIFICVIS.2009.4906848}
{doi: {{%
10\hspace{.1pt}\discretionary{.}{%
}{.}\hspace{.4pt}1109\discretionary{/}{%
}{/}PACIFICVIS\hspace{.1pt}\discretionary{.}{%
}{.}\hspace{.4pt}2009\hspace{.1pt}\discretionary{.}{%
}{.}\hspace{.4pt}4906848}}}


\bibitem{hhe-eca-08}
\href{https://doi.org/10.1109/PACIFICVIS.2008.4475457}{W.~Huang, S.-H. Hong,
  and P.~Eades}.
\newblock \href{https://doi.org/10.1109/PACIFICVIS.2008.4475457}{Effects of
  crossing angles}.
\newblock \href{https://doi.org/10.1109/PACIFICVIS.2008.4475457}{In {\em
  Pacific Visualization Symposium (PacificVis'08)}},
  \href{https://doi.org/10.1109/PACIFICVIS.2008.4475457}{pp. 41--46}.
  \href{https://doi.org/10.1109/PACIFICVIS.2008.4475457}{{IEEE}},
  \href{https://doi.org/10.1109/PACIFICVIS.2008.4475457}{2008}.
  \href{https://doi.org/10.1109/PACIFICVIS.2008.4475457}
{doi: {{%
10\hspace{.1pt}\discretionary{.}{%
}{.}\hspace{.4pt}1109\discretionary{/}{%
}{/}PACIFICVIS\hspace{.1pt}\discretionary{.}{%
}{.}\hspace{.4pt}2008\hspace{.1pt}\discretionary{.}{%
}{.}\hspace{.4pt}4475457}}}


\bibitem{hpss-ttcd-07}
\href{https://doi.org/10.1007/s00453-006-0165-x}{P.~Hui, M.~J. Pelsmajer,
  M.~Schaefer, and D.~{\v S}tefankovi{\v c}}.
\newblock \href{https://doi.org/10.1007/s00453-006-0165-x}{Train tracks and
  confluent drawings}.
\newblock \href{https://doi.org/10.1007/s00453-006-0165-x}{{\em Algorithmica}},
  \href{https://doi.org/10.1007/s00453-006-0165-x}{47:465--479},
  \href{https://doi.org/10.1007/s00453-006-0165-x}{2007}.
  \href{https://doi.org/10.1007/s00453-006-0165-x}
{doi: {{%
10\hspace{.1pt}\discretionary{.}{%
}{.}\hspace{.4pt}1007\discretionary{/}{%
}{/}s00453\discretionary{%
}{-}{-}006\discretionary{%
}{-}{-}0165\discretionary{%
}{-}{-}x}}}


\bibitem{hurter_2012}
\href{https://doi.org/https://doi.org/10.1111/j.1467-8659.2012.03079.x}{C.~Hurter,
  O.~Ersoy, and A.~Telea}.
\newblock
  \href{https://doi.org/https://doi.org/10.1111/j.1467-8659.2012.03079.x}{Graph
  bundling by kernel density estimation}.
\newblock
  \href{https://doi.org/https://doi.org/10.1111/j.1467-8659.2012.03079.x}{{\em
  Computer Graphics Forum}},
  \href{https://doi.org/https://doi.org/10.1111/j.1467-8659.2012.03079.x}{31(3):865--874},
  \href{https://doi.org/https://doi.org/10.1111/j.1467-8659.2012.03079.x}{2012}.
  \href{https://doi.org/10.1111/j.1467-8659.2012.03079.x}
{doi: {{%
10\hspace{.1pt}\discretionary{.}{%
}{.}\hspace{.4pt}1111\discretionary{/}{%
}{/}j\hspace{.1pt}\discretionary{.}{%
}{.}\hspace{.4pt}1467\discretionary{%
}{-}{-}8659\hspace{.1pt}\discretionary{.}{%
}{.}\hspace{.4pt}2012\hspace{.1pt}\discretionary{.}{%
}{.}\hspace{.4pt}03079\hspace{.1pt}\discretionary{.}{%
}{.}\hspace{.4pt}x}}}


\bibitem{johnson_1991}
B.~Johnson and B.~Shneiderman.
\newblock {Tree-Maps}: A space-filling approach to the visualization of
  hierarchical information structures.
\newblock In {\em Proc. of the 2nd Conference on Visualization '91}, VIS '91,
  p. 284–291, 1991.

\bibitem{lambert_2010}
\href{https://doi.org/10.1111/j.1467-8659.2009.01700.x}{A.~Lambert, R.~Bourqui,
  and D.~Auber}.
\newblock \href{https://doi.org/10.1111/j.1467-8659.2009.01700.x}{Winding
  roads: Routing edges into bundles}.
\newblock \href{https://doi.org/10.1111/j.1467-8659.2009.01700.x}{{\em Comput.
  Graph. Forum}},
  \href{https://doi.org/10.1111/j.1467-8659.2009.01700.x}{29(3):853--862},
  \href{https://doi.org/10.1111/j.1467-8659.2009.01700.x}{2010}.
  \href{https://doi.org/10.1111/j.1467-8659.2009.01700.x}
{doi: {{%
10\hspace{.1pt}\discretionary{.}{%
}{.}\hspace{.4pt}1111\discretionary{/}{%
}{/}j\hspace{.1pt}\discretionary{.}{%
}{.}\hspace{.4pt}1467\discretionary{%
}{-}{-}8659\hspace{.1pt}\discretionary{.}{%
}{.}\hspace{.4pt}2009\hspace{.1pt}\discretionary{.}{%
}{.}\hspace{.4pt}01700\hspace{.1pt}\discretionary{.}{%
}{.}\hspace{.4pt}x}}}


\bibitem{lambert_2011}
\href{https://doi.org/https://doi.org/10.1111/j.1467-8659.2011.01951.x}{A.~Lambert,
  J.~Dubois, and R.~Bourqui}.
\newblock
  \href{https://doi.org/https://doi.org/10.1111/j.1467-8659.2011.01951.x}{Pathway
  preserving representation of metabolic networks}.
\newblock
  \href{https://doi.org/https://doi.org/10.1111/j.1467-8659.2011.01951.x}{{\em
  Computer Graphics Forum}},
  \href{https://doi.org/https://doi.org/10.1111/j.1467-8659.2011.01951.x}{30(3):1021--1030},
  \href{https://doi.org/https://doi.org/10.1111/j.1467-8659.2011.01951.x}{2011}.
  \href{https://doi.org/10.1111/j.1467-8659.2011.01951.x}
{doi: {{%
10\hspace{.1pt}\discretionary{.}{%
}{.}\hspace{.4pt}1111\discretionary{/}{%
}{/}j\hspace{.1pt}\discretionary{.}{%
}{.}\hspace{.4pt}1467\discretionary{%
}{-}{-}8659\hspace{.1pt}\discretionary{.}{%
}{.}\hspace{.4pt}2011\hspace{.1pt}\discretionary{.}{%
}{.}\hspace{.4pt}01951\hspace{.1pt}\discretionary{.}{%
}{.}\hspace{.4pt}x}}}


\bibitem{LeskovecAH06}
\href{https://doi.org/10.1145/1134707.1134732}{J.~Leskovec, L.~A. Adamic, and
  B.~A. Huberman}.
\newblock \href{https://doi.org/10.1145/1134707.1134732}{The dynamics of viral
  marketing}.
\newblock \href{https://doi.org/10.1145/1134707.1134732}{In J.~Feigenbaum,
  J.~C. Chuang, and D.~M. Pennock, eds., {\em Proceedings 7th {ACM} Conference
  on Electronic Commerce (EC-2006)}},
  \href{https://doi.org/10.1145/1134707.1134732}{pp. 228--237}.
  \href{https://doi.org/10.1145/1134707.1134732}{{ACM}},
  \href{https://doi.org/10.1145/1134707.1134732}{2006}.
  \href{https://doi.org/10.1145/1134707.1134732}
{doi: {{%
10\hspace{.1pt}\discretionary{.}{%
}{.}\hspace{.4pt}1145\discretionary{/}{%
}{/}1134707\hspace{.1pt}\discretionary{.}{%
}{.}\hspace{.4pt}1134732}}}


\bibitem{snapnets}
J.~Leskovec and A.~Krevl.
\newblock {SNAP Datasets}: {Stanford} large network dataset collection.
\newblock \url{http://snap.stanford.edu/data}, June 2014.
\newblock [Online; accessed 7. Jul. 2021].

\bibitem{lhuiller_2017}
\href{https://doi.org/10.1109/PACIFICVIS.2017.8031594}{A.~{Lhuillier},
  C.~{Hurter}, and A.~{Telea}}.
\newblock \href{https://doi.org/10.1109/PACIFICVIS.2017.8031594}{{FFTEB:} edge
  bundling of huge graphs by the fast fourier transform}.
\newblock \href{https://doi.org/10.1109/PACIFICVIS.2017.8031594}{In {\em
  Pacific Visualization Symposium (PacificVis'17)}},
  \href{https://doi.org/10.1109/PACIFICVIS.2017.8031594}{pp. 190--199},
  \href{https://doi.org/10.1109/PACIFICVIS.2017.8031594}{2017}.
  \href{https://doi.org/10.1109/PACIFICVIS.2017.8031594}
{doi: {{%
10\hspace{.1pt}\discretionary{.}{%
}{.}\hspace{.4pt}1109\discretionary{/}{%
}{/}PACIFICVIS\hspace{.1pt}\discretionary{.}{%
}{.}\hspace{.4pt}2017\hspace{.1pt}\discretionary{.}{%
}{.}\hspace{.4pt}8031594}}}


\bibitem{lhuillier_2017}
\href{https://doi.org/https://doi.org/10.1111/cgf.13213}{A.~Lhuillier,
  C.~Hurter, and A.~Telea}.
\newblock \href{https://doi.org/https://doi.org/10.1111/cgf.13213}{State of the
  art in edge and trail bundling techniques}.
\newblock \href{https://doi.org/https://doi.org/10.1111/cgf.13213}{{\em
  Computer Graphics Forum}},
  \href{https://doi.org/https://doi.org/10.1111/cgf.13213}{36(3):619--645},
  \href{https://doi.org/https://doi.org/10.1111/cgf.13213}{2017}.
  \href{https://doi.org/10.1111/cgf.13213}
{doi: {{%
10\hspace{.1pt}\discretionary{.}{%
}{.}\hspace{.4pt}1111\discretionary{/}{%
}{/}cgf\hspace{.1pt}\discretionary{.}{%
}{.}\hspace{.4pt}13213}}}


\bibitem{luo_2012}
\href{https://doi.org/10.1109/TVCG.2011.104}{S.~{Luo}, C.~{Liu}, B.~{Chen}, and
  K.~{Ma}}.
\newblock \href{https://doi.org/10.1109/TVCG.2011.104}{Ambiguity-free
  edge-bundling for interactive graph visualization}.
\newblock \href{https://doi.org/10.1109/TVCG.2011.104}{{\em IEEE Trans.
  Visualization and Computer Graphics}},
  \href{https://doi.org/10.1109/TVCG.2011.104}{18(5):810--821},
  \href{https://doi.org/10.1109/TVCG.2011.104}{2012}.
  \href{https://doi.org/10.1109/TVCG.2011.104}
{doi: {{%
10\hspace{.1pt}\discretionary{.}{%
}{.}\hspace{.4pt}1109\discretionary{/}{%
}{/}TVCG\hspace{.1pt}\discretionary{.}{%
}{.}\hspace{.4pt}2011\hspace{.1pt}\discretionary{.}{%
}{.}\hspace{.4pt}104}}}


\bibitem{neh_ofgv_2013}
\href{https://doi.org/10.1109/PacificVis.2013.6596147}{Q.~H. Nguyen, P.~Eades,
  and S.~Hong}.
\newblock \href{https://doi.org/10.1109/PacificVis.2013.6596147}{On the
  {F}aithfulness of {G}raph {V}isualizations}.
\newblock \href{https://doi.org/10.1109/PacificVis.2013.6596147}{In {\em
  Pacific Visualization Symposium (PacificVis'13)}},
  \href{https://doi.org/10.1109/PacificVis.2013.6596147}{pp. 209--216}.
  \href{https://doi.org/10.1109/PacificVis.2013.6596147}{{IEEE}},
  \href{https://doi.org/10.1109/PacificVis.2013.6596147}{2013}.
  \href{https://doi.org/10.1109/PacificVis.2013.6596147}
{doi: {{%
10\hspace{.1pt}\discretionary{.}{%
}{.}\hspace{.4pt}1109\discretionary{/}{%
}{/}PacificVis\hspace{.1pt}\discretionary{.}{%
}{.}\hspace{.4pt}2013\hspace{.1pt}\discretionary{.}{%
}{.}\hspace{.4pt}6596147}}}


\bibitem{NguyenEH17}
\href{http://arxiv.org/abs/1701.00921}{Q.~H. Nguyen, P.~Eades, and S.~Hong}.
\newblock \href{http://arxiv.org/abs/1701.00921}{Towards faithful graph
  visualizations}.
\newblock \href{http://arxiv.org/abs/1701.00921}{{\em CoRR}},
  \href{http://arxiv.org/abs/1701.00921}{abs/1701.00921},
  \href{http://arxiv.org/abs/1701.00921}{2017}.

\bibitem{nguyen_2012}
\href{https://doi.org/10.1007/978-3-642-25878-7_13}{Q.~H. Nguyen, S.~Hong, and
  P.~Eades}.
\newblock \href{https://doi.org/10.1007/978-3-642-25878-7_13}{{TGI-EB:} {A} new
  framework for edge bundling integrating topology, geometry and importance}.
\newblock \href{https://doi.org/10.1007/978-3-642-25878-7_13}{In M.~J. van
  Kreveld and B.~Speckmann, eds., {\em Graph Drawing (GD'11)}},
  \href{https://doi.org/10.1007/978-3-642-25878-7_13}{vol. 7034 of {\em LNCS}},
  \href{https://doi.org/10.1007/978-3-642-25878-7_13}{pp. 123--135}.
  \href{https://doi.org/10.1007/978-3-642-25878-7_13}{Springer},
  \href{https://doi.org/10.1007/978-3-642-25878-7_13}{2011}.
  \href{https://doi.org/10.1007/978-3-642-25878-7_13}
{doi: {{%
10\hspace{.1pt}\discretionary{.}{%
}{.}\hspace{.4pt}1007\discretionary{/}{%
}{/}978\discretionary{%
}{-}{-}3\discretionary{%
}{-}{-}642\discretionary{%
}{-}{-}25878\discretionary{%
}{-}{-}7\_13}}}


\bibitem{palmas_2014}
\href{https://doi.org/10.1109/PacificVis.2014.40}{G.~{Palmas}, M.~{Bachynskyi},
  A.~{Oulasvirta}, H.~P. {Seidel}, and T.~{Weinkauf}}.
\newblock \href{https://doi.org/10.1109/PacificVis.2014.40}{An edge-bundling
  layout for interactive parallel coordinates}.
\newblock \href{https://doi.org/10.1109/PacificVis.2014.40}{In {\em Pacific
  Visualization Symposium (PacificVis'14)}},
  \href{https://doi.org/10.1109/PacificVis.2014.40}{pp. 57--64}.
  \href{https://doi.org/10.1109/PacificVis.2014.40}{IEEE},
  \href{https://doi.org/10.1109/PacificVis.2014.40}{2014}.
  \href{https://doi.org/10.1109/PacificVis.2014.40}
{doi: {{%
10\hspace{.1pt}\discretionary{.}{%
}{.}\hspace{.4pt}1109\discretionary{/}{%
}{/}PacificVis\hspace{.1pt}\discretionary{.}{%
}{.}\hspace{.4pt}2014\hspace{.1pt}\discretionary{.}{%
}{.}\hspace{.4pt}40}}}


\bibitem{phan_2005}
\href{https://doi.org/10.1109/INFVIS.2005.1532150}{D.~Phan, L.~Xiao, R.~B. Yeh,
  P.~Hanrahan, and T.~Winograd}.
\newblock \href{https://doi.org/10.1109/INFVIS.2005.1532150}{Flow map layout}.
\newblock \href{https://doi.org/10.1109/INFVIS.2005.1532150}{In J.~T. Stasko
  and M.~O. Ward, eds., {\em Information Visualization (InfoVis'05)}},
  \href{https://doi.org/10.1109/INFVIS.2005.1532150}{pp. 219--224}.
  \href{https://doi.org/10.1109/INFVIS.2005.1532150}{{IEEE}},
  \href{https://doi.org/10.1109/INFVIS.2005.1532150}{2005}.
  \href{https://doi.org/10.1109/INFVIS.2005.1532150}
{doi: {{%
10\hspace{.1pt}\discretionary{.}{%
}{.}\hspace{.4pt}1109\discretionary{/}{%
}{/}INFVIS\hspace{.1pt}\discretionary{.}{%
}{.}\hspace{.4pt}2005\hspace{.1pt}\discretionary{.}{%
}{.}\hspace{.4pt}1532150}}}


\bibitem{puprev_2016}
\href{https://doi.org/https://doi.org/10.1016/j.comgeo.2015.10.005}{S.~Pupyrev,
  L.~Nachmanson, S.~Bereg, and A.~E. Holroyd}.
\newblock
  \href{https://doi.org/https://doi.org/10.1016/j.comgeo.2015.10.005}{Edge
  routing with ordered bundles}.
\newblock
  \href{https://doi.org/https://doi.org/10.1016/j.comgeo.2015.10.005}{{\em
  Computational Geometry}},
  \href{https://doi.org/https://doi.org/10.1016/j.comgeo.2015.10.005}{52:18--33},
  \href{https://doi.org/https://doi.org/10.1016/j.comgeo.2015.10.005}{2016}.
  \href{https://doi.org/10.1016/j.comgeo.2015.10.005}
{doi: {{%
10\hspace{.1pt}\discretionary{.}{%
}{.}\hspace{.4pt}1016\discretionary{/}{%
}{/}j\hspace{.1pt}\discretionary{.}{%
}{.}\hspace{.4pt}comgeo\hspace{.1pt}\discretionary{.}{%
}{.}\hspace{.4pt}2015\hspace{.1pt}\discretionary{.}{%
}{.}\hspace{.4pt}10\hspace{.1pt}\discretionary{.}{%
}{.}\hspace{.4pt}005}}}


\bibitem{reniers_2014}
\href{https://doi.org/https://doi.org/10.1016/j.scico.2012.05.002}{D.~Reniers,
  L.~Voinea, O.~Ersoy, and A.~Telea}.
\newblock
  \href{https://doi.org/https://doi.org/10.1016/j.scico.2012.05.002}{The solid*
  toolset for software visual analytics of program structure and metrics
  comprehension: From research prototype to product}.
\newblock
  \href{https://doi.org/https://doi.org/10.1016/j.scico.2012.05.002}{{\em
  Science of Computer Programming}},
  \href{https://doi.org/https://doi.org/10.1016/j.scico.2012.05.002}{79:224--240},
  \href{https://doi.org/https://doi.org/10.1016/j.scico.2012.05.002}{2014}.
  \href{https://doi.org/10.1016/j.scico.2012.05.002}
{doi: {{%
10\hspace{.1pt}\discretionary{.}{%
}{.}\hspace{.4pt}1016\discretionary{/}{%
}{/}j\hspace{.1pt}\discretionary{.}{%
}{.}\hspace{.4pt}scico\hspace{.1pt}\discretionary{.}{%
}{.}\hspace{.4pt}2012\hspace{.1pt}\discretionary{.}{%
}{.}\hspace{.4pt}05\hspace{.1pt}\discretionary{.}{%
}{.}\hspace{.4pt}002}}}


\bibitem{selassie_2011}
\href{https://doi.org/10.1109/TVCG.2011.190}{D.~{Selassie}, B.~{Heller}, and
  J.~{Heer}}.
\newblock \href{https://doi.org/10.1109/TVCG.2011.190}{Divided edge bundling
  for directional network data}.
\newblock \href{https://doi.org/10.1109/TVCG.2011.190}{{\em IEEE Trans.
  Visualization and Computer Graphics}},
  \href{https://doi.org/10.1109/TVCG.2011.190}{17(12):2354--2363},
  \href{https://doi.org/10.1109/TVCG.2011.190}{2011}.
  \href{https://doi.org/10.1109/TVCG.2011.190}
{doi: {{%
10\hspace{.1pt}\discretionary{.}{%
}{.}\hspace{.4pt}1109\discretionary{/}{%
}{/}TVCG\hspace{.1pt}\discretionary{.}{%
}{.}\hspace{.4pt}2011\hspace{.1pt}\discretionary{.}{%
}{.}\hspace{.4pt}190}}}


\bibitem{telea_2010}
\href{https://doi.org/https://doi.org/10.1111/j.1467-8659.2009.01680.x}{A.~Telea
  and O.~Ersoy}.
\newblock
  \href{https://doi.org/https://doi.org/10.1111/j.1467-8659.2009.01680.x}{Image-based
  edge bundles: Simplified visualization of large graphs}.
\newblock
  \href{https://doi.org/https://doi.org/10.1111/j.1467-8659.2009.01680.x}{{\em
  Computer Graphics Forum}},
  \href{https://doi.org/https://doi.org/10.1111/j.1467-8659.2009.01680.x}{29(3):843--852},
  \href{https://doi.org/https://doi.org/10.1111/j.1467-8659.2009.01680.x}{2010}.
  \href{https://doi.org/10.1111/j.1467-8659.2009.01680.x}
{doi: {{%
10\hspace{.1pt}\discretionary{.}{%
}{.}\hspace{.4pt}1111\discretionary{/}{%
}{/}j\hspace{.1pt}\discretionary{.}{%
}{.}\hspace{.4pt}1467\discretionary{%
}{-}{-}8659\hspace{.1pt}\discretionary{.}{%
}{.}\hspace{.4pt}2009\hspace{.1pt}\discretionary{.}{%
}{.}\hspace{.4pt}01680\hspace{.1pt}\discretionary{.}{%
}{.}\hspace{.4pt}x}}}


\bibitem{toeda_2017}
\href{https://doi.org/https://doi.org/10.1016/j.jvlc.2017.09.004}{N.~Toeda,
  R.~Nakazawa, T.~Itoh, T.~Saito, and D.~Archambault}.
\newblock
  \href{https://doi.org/https://doi.org/10.1016/j.jvlc.2017.09.004}{Convergent
  drawing for mutually connected directed graphs}.
\newblock
  \href{https://doi.org/https://doi.org/10.1016/j.jvlc.2017.09.004}{{\em J.
  Visual Languages \& Computing}},
  \href{https://doi.org/https://doi.org/10.1016/j.jvlc.2017.09.004}{43:83--90},
  \href{https://doi.org/https://doi.org/10.1016/j.jvlc.2017.09.004}{2017}.
  \href{https://doi.org/10.1016/j.jvlc.2017.09.004}
{doi: {{%
10\hspace{.1pt}\discretionary{.}{%
}{.}\hspace{.4pt}1016\discretionary{/}{%
}{/}j\hspace{.1pt}\discretionary{.}{%
}{.}\hspace{.4pt}jvlc\hspace{.1pt}\discretionary{.}{%
}{.}\hspace{.4pt}2017\hspace{.1pt}\discretionary{.}{%
}{.}\hspace{.4pt}09\hspace{.1pt}\discretionary{.}{%
}{.}\hspace{.4pt}004}}}


\bibitem{t-vdqi-83}
E.~R. Tufte.
\newblock {\em The Visual Display of Quantitative Information}.
\newblock Graphics Press, 1983.

\bibitem{migrations}
\href{https://www.census.gov/data/tables/2000/demo/geographic-mobility/county-to-county-migration-flows.html}{{United
  States Census Bureau}}.
\newblock
  \href{https://www.census.gov/data/tables/2000/demo/geographic-mobility/county-to-county-migration-flows.html}{County-to-county
  migration flows files: Census 2000},
  \href{https://www.census.gov/data/tables/2000/demo/geographic-mobility/county-to-county-migration-flows.html}{2000}.
\newblock
  \href{https://www.census.gov/data/tables/2000/demo/geographic-mobility/county-to-county-migration-flows.html}{[Online;
  accessed 7. Jul. 2021]}.

\bibitem{vdzwan_2016}
\href{https://doi.org/10.1109/TVCG.2016.2515611}{M.~{van der Zwan},
  V.~{Codreanu}, and A.~{Telea}}.
\newblock \href{https://doi.org/10.1109/TVCG.2016.2515611}{{CUBu:} universal
  real-time bundling for large graphs}.
\newblock \href{https://doi.org/10.1109/TVCG.2016.2515611}{{\em IEEE Trans.
  Visualization and Computer Graphics}},
  \href{https://doi.org/10.1109/TVCG.2016.2515611}{22(12):2550--2563},
  \href{https://doi.org/10.1109/TVCG.2016.2515611}{2016}.
  \href{https://doi.org/10.1109/TVCG.2016.2515611}
{doi: {{%
10\hspace{.1pt}\discretionary{.}{%
}{.}\hspace{.4pt}1109\discretionary{/}{%
}{/}TVCG\hspace{.1pt}\discretionary{.}{%
}{.}\hspace{.4pt}2016\hspace{.1pt}\discretionary{.}{%
}{.}\hspace{.4pt}2515611}}}


\bibitem{vanHam_2004}
\href{https://doi.org/10.1109/INFVIS.2004.43}{F.~{van Ham} and J.~J. {van
  Wijk}}.
\newblock \href{https://doi.org/10.1109/INFVIS.2004.43}{Interactive
  visualization of small world graphs}.
\newblock \href{https://doi.org/10.1109/INFVIS.2004.43}{In {\em IEEE Symposium
  on Information Visualization}},
  \href{https://doi.org/10.1109/INFVIS.2004.43}{pp. 199--206},
  \href{https://doi.org/10.1109/INFVIS.2004.43}{2004}.
  \href{https://doi.org/10.1109/INFVIS.2004.43}
{doi: {{%
10\hspace{.1pt}\discretionary{.}{%
}{.}\hspace{.4pt}1109\discretionary{/}{%
}{/}INFVIS\hspace{.1pt}\discretionary{.}{%
}{.}\hspace{.4pt}2004\hspace{.1pt}\discretionary{.}{%
}{.}\hspace{.4pt}43}}}


\bibitem{wang_2016}
\href{https://doi.org/10.1109/TVCG.2015.2467691}{Y.~{Wang}, Q.~{Shen},
  D.~{Archambault}, Z.~{Zhou}, M.~{Zhu}, S.~{Yang}, and H.~{Qu}}.
\newblock \href{https://doi.org/10.1109/TVCG.2015.2467691}{{AmbiguityVis:}
  visualization of ambiguity in graph layouts}.
\newblock \href{https://doi.org/10.1109/TVCG.2015.2467691}{{\em IEEE Trans.
  Visualization and Computer Graphics}},
  \href{https://doi.org/10.1109/TVCG.2015.2467691}{22(1):359--368},
  \href{https://doi.org/10.1109/TVCG.2015.2467691}{2016}.
  \href{https://doi.org/10.1109/TVCG.2015.2467691}
{doi: {{%
10\hspace{.1pt}\discretionary{.}{%
}{.}\hspace{.4pt}1109\discretionary{/}{%
}{/}TVCG\hspace{.1pt}\discretionary{.}{%
}{.}\hspace{.4pt}2015\hspace{.1pt}\discretionary{.}{%
}{.}\hspace{.4pt}2467691}}}


\bibitem{wong_2003}
\href{https://doi.org/10.1109/INFVIS.2003.1249008}{N.~Wong, S.~Carpendale, and
  S.~Greenberg}.
\newblock \href{https://doi.org/10.1109/INFVIS.2003.1249008}{Edgelens: An
  interactive method for managing edge congestion in graphs}.
\newblock \href{https://doi.org/10.1109/INFVIS.2003.1249008}{In {\em Symposium
  on Information Visualization (InfoVis'03)}},
  \href{https://doi.org/10.1109/INFVIS.2003.1249008}{pp. 51--58}.
  \href{https://doi.org/10.1109/INFVIS.2003.1249008}{{IEEE}},
  \href{https://doi.org/10.1109/INFVIS.2003.1249008}{2003}.
  \href{https://doi.org/10.1109/INFVIS.2003.1249008}
{doi: {{%
10\hspace{.1pt}\discretionary{.}{%
}{.}\hspace{.4pt}1109\discretionary{/}{%
}{/}INFVIS\hspace{.1pt}\discretionary{.}{%
}{.}\hspace{.4pt}2003\hspace{.1pt}\discretionary{.}{%
}{.}\hspace{.4pt}1249008}}}


\bibitem{wu_2015}
\href{https://doi.org/10.1109/BigData.2015.7364046}{J.~{Wu}, L.~{Yu}, and
  H.~{Yu}}.
\newblock \href{https://doi.org/10.1109/BigData.2015.7364046}{Texture-based
  edge bundling: A web-based approach for interactively visualizing large
  graphs}.
\newblock \href{https://doi.org/10.1109/BigData.2015.7364046}{In {\em Big
  Data}}, \href{https://doi.org/10.1109/BigData.2015.7364046}{pp. 2501--2508}.
  \href{https://doi.org/10.1109/BigData.2015.7364046}{IEEE},
  \href{https://doi.org/10.1109/BigData.2015.7364046}{2015}.
  \href{https://doi.org/10.1109/BigData.2015.7364046}
{doi: {{%
10\hspace{.1pt}\discretionary{.}{%
}{.}\hspace{.4pt}1109\discretionary{/}{%
}{/}BigData\hspace{.1pt}\discretionary{.}{%
}{.}\hspace{.4pt}2015\hspace{.1pt}\discretionary{.}{%
}{.}\hspace{.4pt}7364046}}}


\bibitem{WuZhuLiuYu18}
\href{https://doi.org/10.3390/e20090625}{J.~Wu, F.~Zhu, X.~Liu, and H.~Yu}.
\newblock \href{https://doi.org/10.3390/e20090625}{An information-theoretic
  framework for evaluating edge bundling visualization}.
\newblock \href{https://doi.org/10.3390/e20090625}{{\em Entropy}},
  \href{https://doi.org/10.3390/e20090625}{20(9):625},
  \href{https://doi.org/10.3390/e20090625}{2018}.
  \href{https://doi.org/10.3390/e20090625}
{doi: {{%
10\hspace{.1pt}\discretionary{.}{%
}{.}\hspace{.4pt}3390\discretionary{/}{%
}{/}e20090625}}}


\bibitem{ZhangLGJHLH20}
\href{https://doi.org/10.1145/3416495}{Y.~Zhang, X.~Liao, L.~Gu, H.~Jin, K.~Hu,
  H.~Liu, and B.~He}.
\newblock \href{https://doi.org/10.1145/3416495}{{AsynGraph}: Maximizing data
  parallelism for efficient iterative graph processing on gpus}.
\newblock \href{https://doi.org/10.1145/3416495}{{\em {ACM} Trans. Archit. Code
  Optim.}}, \href{https://doi.org/10.1145/3416495}{17(4):29:1--29:21},
  \href{https://doi.org/10.1145/3416495}{2020}.
  \href{https://doi.org/10.1145/3416495}
{doi: {{%
10\hspace{.1pt}\discretionary{.}{%
}{.}\hspace{.4pt}1145\discretionary{/}{%
}{/}3416495}}}


\bibitem{zheng_2021}
\href{https://doi.org/10.1109/TVCG.2019.2944619}{J.~X. {Zheng}, S.~{Pawar}, and
  D.~F.~M. {Goodman}}.
\newblock \href{https://doi.org/10.1109/TVCG.2019.2944619}{Further towards
  unambiguous edge bundling: Investigating power-confluent drawings for network
  visualization}.
\newblock \href{https://doi.org/10.1109/TVCG.2019.2944619}{{\em IEEE Trans.
  Visualization and Computer Graphics}},
  \href{https://doi.org/10.1109/TVCG.2019.2944619}{27(3):2244--2249},
  \href{https://doi.org/10.1109/TVCG.2019.2944619}{2021}.
  \href{https://doi.org/10.1109/TVCG.2019.2944619}
{doi: {{%
10\hspace{.1pt}\discretionary{.}{%
}{.}\hspace{.4pt}1109\discretionary{/}{%
}{/}TVCG\hspace{.1pt}\discretionary{.}{%
}{.}\hspace{.4pt}2019\hspace{.1pt}\discretionary{.}{%
}{.}\hspace{.4pt}2944619}}}


\end{thebibliography}

\end{document}